\documentclass[reprint,amsmath,twocolumn,amssymb, nobibnotes, aps, pre, superscriptaddress]{revtex4-1}

\usepackage{graphicx}
\usepackage{dcolumn}
\usepackage{bm}
\usepackage{color}
\usepackage{tikz}
\usetikzlibrary{shapes}
\usetikzlibrary{backgrounds}

\begin{document}
\def \beq{\begin{equation}}
\def \eeq{\end{equation}}
\def \bea{\begin{eqnarray}}
\def \eea{\end{eqnarray}}
\def \bem{\begin{displaymath}}
\def \eem{\end{displaymath}}
\def \P{\Psi}
\def \Pd{|\Psi(\boldsymbol{r})|}
\def \Pds{|\Psi^{\ast}(\boldsymbol{r})|}
\def \Po{\overline{\Psi}}
\def \bs{\boldsymbol}
\def \bl{\bar{\boldsymbol{l}}}

\title{ {Tight{-}binding methods for general longitudinally driven {photonic lattices --} edge states and solitons}}
\author{Mark J. Ablowitz and Justin T. Cole}
\affiliation{Department of Applied Mathematics, University of Colorado, Boulder, Colorado 80309}
\date{\today}     
\begin{abstract}

A systematic approach for deriving {tight-binding approximations}  in  {general longitudinally} driven lattices is presented.  {As prototypes,} honeycomb and staggered square lattices are considered. Time-reversal symmetry is broken by 
{varying/rotating} the waveguides, {longitudinally,} along the direction of propagation. {Different sublattice rotation and structure are allowed.}
Linear Floquet bands are constructed for intricate sublattice rotation patterns such as counter rotation, phase offset rotation, {as well as different lattice sizes and frequencies.} 
An asymptotic analysis of the edge modes, valid in a rapid-spiraling regime, reveals linear and nonlinear envelopes which are governed by linear and nonlinear Schr\"odinger equations, respectively.
{Nonlinear states, referred to as topologically protected edge solitons}{ are unidirectional edge modes.}
{Direct numerical simulations for both the linear and nonlinear edge states agree} with the asymptotic theory. 
{Topologically} protected modes are {found; they} possess unidirectionality and do not scatter at lattice defect boundaries. 

\begin{description}

\item[PACS numbers] 42.65.Tg, 42.65.Jx, 42.82.Et
\end{description}
\end{abstract}
\pacs{Valid PACS appear here}
\maketitle

\section{Introduction}
\label{intro}
{In recent years the study of topological edge/interface and surface modes has been a very active area of research.}
{
Fundamental research in systems where time-reversal symmetry is broken have been found to support unidirectional edge modes that remain intact over long distance and against defects; these systems include magneto-optics \cite{Wang2008, Wang2009}, photonics \cite{HaldRag2008, Hafezi2013, RechtsSegev2013} and acoustics \cite{ZYang2015}.}  {While} these systems have very different underlying physics they all share {some} common properties{; they have a} boundary that separates two distinct regions. A boundary discontinuity alone can be enough to support edge states, but to generate topologically protected modes in 
 {these systems typically  time-reversal symmetry 
must be broken.} The manner in which this is accomplished is different for each system{; here} however we focus on optical beams propagating through {longitudinally} driven waveguide arrays.

{Some of the} recent advances in topologically protected systems include: PT symmetry crystals \cite{Weimann2016}, bi-anisotropic metamaterials \cite{Khanikaev2013, Chen2014} and surface plasmons \cite{DiPietro2013}.  {Higher-dimensional systems have also} been examined \cite{Lu2012}, where the 3D analog of Dirac points called Weyl points {have been} {reported} \cite{Wan2011, Noh2017}.
{Quasi-crystal type arrays can also support} unidirectional edge states \cite{Bandres2016}.

In photonic lattices, regions of high refractive index {can be} etched into bulk silica using {f}emtosecond laser etching technique{s} \cite{Szameit2006, Szameit2006A} and act as waveguides for the beam. One way to break time-reversal symmetry is to rotate these waveguides {longitudinally,  creating  an  array configuration that changes along along the direction of beam propagation.} {Such lattices, which are orthogonal to the direction of the beam propagation,} have been constructed at optical frequencies and topologically protected edge modes have been experimentally observed \cite{RechtsSegev2013}.

One of the {special features  topological edge modes exhibit} is their one-way scatter-free motion even when lattice defects are encountered.
{Such traveling modes are said to be {\it topologically protected} meaning that unidirectionality is preserved even under significant deformations of} the lattice. Topological invariants, such as Chern number \cite{Rudner2013} or Zak phase \cite{Zak1989}, {can be associated with these topological modes.}
The robust nature of topologically protected edge modes {suggests that they will be useful in many applications where small imperfections are always present.}


{Photonic lattices  that can support linear edge modes include honeycomb \cite{RechtsSegev2013} and staggered square lattices \cite{Leykam2016,Leykam2016A} each of  which has two lattice} sites per unit cell. We refer to any lattice with two or more sites in a unit cell as {\it non-simple}. 
In this work we provide a direct route for deriving tight-binding equations that describe beam propagation in {general longitudinally varying lattices with {either simple or } non-simple configurations. As typical examples we analyze honeycomb and staggered square lattices, though more complex lattices can be considered within the framework we present.}  In doing so, we are able to {study
a wide range of lattice dynamics including periodic but non-synchronized (out of phase) waveguide motion with phase offset across} the two sublattices or counter rotating lattices etc. The { Floquet bands are associated with 
 intriguing} edge mode dynamics which include: flat (stationary), {non-unidirectional, oscillatory, and simultaneous topological and non-topological eigenstates}.
These modes complement the well-known traveling edge states present when the lattice waveguides rotate in-phase with each other. 

{For a particular rotation pattern we find common structure in the dispersion bands of both the honeycomb and staggered square lattices. This suggests that for a particular waveguide rotation there is an associated edge wave dynamic that is  {\it independent} of the underlying lattice {configuration.} Additionally, our tight-binding model incorporates the geometry of the individual waveguides.
When the waveguides are stretched in a preferred direction we find non-topologically protected modes whose edge modes reflect at lattice defects, rather than simply moving around the defect like a protected {mode.}}

We 
present an asymptotic analysis which reveals that along the lattice boundary the edge modes {can} behave as one-dimensional solutions to the nonlinear Schr\"odinger (NLS) equation. Hence we refer to these traveling modes as {\it edge solitons}, where applicable. 
The NLS equation is a universal model for {the envelope description of} dispersive waves in weakly-nonlinear media, like the {systems} we consider in this paper. The nonlinear edge mode envelope can be seen as {a} balance between the lattice induced linear dispersion effects and {sufficiently strong} beam focusing or defocusing. Indeed, many  {properties associated with} the NLS equation can be found in photonic topological insulators, such as modulational instability \cite{Lumer2016}, band gap solitons \cite{Lumer2013} {and soliton propagation in helically driven photonic graphene \cite{AbCuMa2014}}. {Moreover, we construct nonlinear modes that, to leading order, satisfy the focusing NLS equation and whose corresponding linear dispersion is topologically protected.} {These modes travel stably around defects.}  {Hence we term these solutions {\it topologically protected edge solitons}. }

{The outline of the paper is as follows.
In} Sec.~\ref{TBA_derive} we derive a tight-binding model that describes paraxial beams  {in general longitudinally} driven honeycomb and staggered square lattices. 
{The} linear dispersion bands for one-dimensional edge states are computed via Floquet theory in Sec.~\ref{linear_theory}. {Here we} study several complex sublattice rotation patterns and their {effect} on the edge mode dynamics. In Sec.~\ref{Nonlinear_modes} we consider {nonlinear} edge beams. Both linear and nonlinear edge modes are found to be modulated by a slowly-varying envelope function that satisfy the linear and nonlinear Schr\"odinger equations, respectively.
In the nonlinear regime {an asymptotic} analysis shows {that} edge solitons exist for solutions with narrow spectral support {around some central frequency.} 
The edge wave behavior {across} lattice defects is explored in Sec.~\ref{defect_sec}.  There we observe topologically protected modes to be scatter-free at lattice barriers, in contrast to non-topologically protected modes.
{We} conclude in Sec.~\ref{conclude}.


\section{Derivation of Tight-binding Approximation}
\label{TBA_derive}
{Nonlinear quasi}-monochromatic light beams propagating through photonic lattices are described by the paraxial wave or nonlinear Schr\"odinger (NLS) equation
\begin{equation}
\label{NLS}
i \frac{ \partial \psi}{\partial z} + \frac{1}{2 k_0} \nabla^2 \psi - \frac{k_0}{n_0}  \left( \Delta n({\bf r},z) + n_2 |\psi|^2 \right)\psi = 0  \; ,
\end{equation}
where $\nabla^2 \equiv \partial_x^2 + \partial_y^2$ and $\psi({\bf r},z)$ is a complex envelope function defined on the transverse plane ${\bf r} = (x,y)$ and propagation direction $z$. The function $\Delta n({\bf r},z)$ is the lattice potential and it models the variations in the refractive index.
The physical parameters are the wavenumber $k_0 = 2 \pi n_0 / \lambda$, bulk index of refraction $n_0$, beam wavelength $\lambda$, deviation from bulk index $|\Delta n|,$ and Kerr coefficient $n_2.$ 

\begin{figure}
\vspace{3mm}
\begin{tikzpicture}[node distance=1cm]

\begin{scope}[xshift=-3cm,yshift=1.5cm]
\node[regular polygon, regular polygon sides=4, shape aspect=0.5, minimum width=3cm, minimum height=1cm, draw,scale=1, rotate=45] (reg) {}; 
 \end{scope}

 \begin{scope}[xshift=-3cm,yshift=-1.5cm]
\node[regular polygon, regular polygon sides=4, shape aspect=0.5, minimum width=3cm, minimum height=1cm, draw,scale=1, rotate=45] (reg) {}; 
 \end{scope}
 
 \begin{scope}[xshift=-1.5cm,yshift=0cm]
\node[regular polygon, regular polygon sides=4, shape aspect=0.5, minimum width=3cm, minimum height=1cm, draw,scale=1, rotate=45] (reg) {}; 
 \end{scope}



 \draw (-1.5,1.5) -- (0,3);
 \draw (-1.5,-1.5) -- (0,-3);
 \draw (0,3) -- (1.5/2,1.5/2+1.5);
 \draw (0,-3) -- (1.5/2,-1.5/2-1.5); 
 \draw (0,0) -- (1.5/2,1.5/2);
 \draw (0,0) -- (1.5/2,-1.5/2);
 \draw (0,3) -- (1.5/2,1.5/2+3);
 \draw (0,-3) -- (1.5/2,-1.5/2-3);  
 \draw (0,3) -- (-1.5/2,1.5/2+3);
 \draw (0,-3) -- (-1.5/2,-1.5/2-3);
 \draw (-3,3) -- (-3+1.5/2,1.5/2+3); 
 \draw (-3,-3) -- (-3+1.5/2,-1.5/2-3);
 \draw (-3,3) -- (-3-1.5/2,1.5/2+3); 
  \draw (-3,-3) -- (-3-1.5/2,-1.5/2-3); 
  
 \fill (-3,0)  circle (2pt);
 \fill (-3,3)  circle (2pt);
 \fill (-3,-3)  circle (2pt);
 \fill (0,0)  circle (2pt);
 \fill (0,3)  circle (2pt);
 \fill (0,-3)  circle (2pt);
 
 \filldraw[fill=white] (-4.5,1.5) circle (2pt);
 \filldraw[fill=white] (-4.5,-1.5) circle (2pt);
 \filldraw[fill=white] (-1.5,-1.5) circle (2pt);
 \filldraw[fill=white] (-1.5,1.5) circle (2pt);

 \draw [line width=0.3mm,->] (-3,0) -- (-1.6,1.4);
 \draw [line width=0.3mm,->] (-3,0) -- (-1.6,-1.4);
 
\draw (-2.5,-0.9) node () { $v_2$};
\draw (-2.5,0.9) node () { $v_1$};
 
 \draw (-0.8,0) node () { $a_{\rm 2m-1,3}$};
 \draw (-3.8,0) node () { $a_{\rm 2m-1,1}$};
 \draw (-0.8,-3) node () { $a_{\rm 2m-3,3}$};
 \draw (-0.8,3) node () { $a_{\rm 2m+1,3}$};
 \draw (-4,-3) node () { $a_{\rm 2m-3,1}$}; 
 \draw (-4,3) node () { $a_{\rm 2m+1,1}$}; 
 
 \draw (-0.7,1.5) node () { $b_{\rm 2m,2}$};
 \draw (-0.7,-1.5) node () { $b_{\rm 2m-2,2}$};
  \draw (-3.7,1.5) node () { $b_{\rm 2m,0}$};
  \draw (-3.7,-1.5) node () { $b_{\rm 2m-2,0}$};

\end{tikzpicture}

 \caption{{The staggered} square lattice consists of two interpenetrating square sublattices $V_1({\bf r})$ and $V_2({\bf r})$ with minima (zeros) located at the white circles and black circles, respectively. The defining lattice vectors are given in Eq.~(\ref{square_lattice_vec_define}).   \label{center_sq_lattice}}

\end{figure}
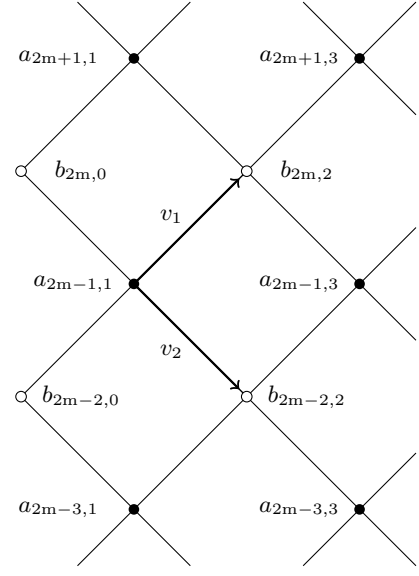

We may think of a lattice with two generating sites per unit cell as the combination of two interpenetrating sublattices {with potentials} $V_1({\bf r})$ and $V_2 ({\bf r})$. These two sublattices have regions of high refractive index concentrated at the white and black lattice sites, respectively, as shown in Figs.~\ref{honeycomb_fig} and \ref{center_sq_lattice}. 
To model such a lattice configuration we {employ} the potential function
\begin{equation}
\label{POT_define}
\Delta n ({\bf r}) =  |\Delta n| \left[ 1 - V_1({\bf r} ) -  V_2({\bf r}) \right] \; ,
\end{equation}
that consists of well-localized dips (minima) located at the lattice sites.
We approximate these sublattices by the sum of Gaussians
\begin{align}
\label{sublattice_define}
& V_{1}({\bf r} ) = \sum_{{\bf r}_j \in \mathcal{W}} \tilde{V}({\bf r} - {\bf r}_j) \; , ~~~
V_{2}({\bf r} ) = \sum_{{\bf r}_k \in \mathcal{B}} \tilde{V}({\bf r} - {\bf r}_k) \; , \\ \nonumber
&~~~~~~~~~ \tilde{V}({\bf r}) =  \exp{\left( - \frac{ x^2 }{\sigma_x^2} - \frac{y^2}{\sigma_y^2} \right)} \; ~~~~  \sigma_x, \sigma_y > 0 \; ,
\end{align}
where the positions of the white and black lattice sites are given, respectively, by the sets
\begin{equation*}
 \mathcal{W} = \left\{ {\bf r}_j : V_1({\bf r}_j) = 1 \right\}  ~ , ~~~
 \mathcal{B} = \left\{ {\bf r}_k : V_2({\bf r}_k) = 1  \right\} .
\end{equation*}
Asymmetry in the waveguide geometry can be explored by adjusting the width parameters $\sigma_x$ and $\sigma_y.$
When $\sigma_x = \sigma_y$ we say the lattice {\it isotropic}, whereas when $\sigma_x \not= \sigma_y$ the lattice is {\it anisotropic}. {Detuning (difference in the refractive index) between the sublattices can approximated by including a coefficient $0 < q < 1$ in front of one of the sublattice potentials i.e. $V_2({\bf r}) \rightarrow q V_2({\bf r})$. {Investigations of} detuned lattices is beyond the scope of this paper and we do not consider it here.} We point out that this approach can be {readily} extended to lattices with more than two lattice sites per unit cell 
by 
{adding} an additional sublattice $V_3({\bf r})$ function to potential function (\ref{POT_define}) that is defined similar to Eq.~(\ref{sublattice_define}).

For well-separated lattice sites, and when ${\bf r}$ is near ${\bf r}_j$ or ${\bf r}_k$, the potential function (\ref{POT_define}) is approximated by the first few terms of its Taylor series and thus takes a paraboloid form, namely
\begin{equation}
\label{paraboloid_pot}
\Delta n ({\bf r}) \approx    |\Delta n| \left( \frac{x^2}{\sigma_x^2} + \frac{y^2}{\sigma_y^2}  \right) \; , ~{\rm as~} {\bf r} \rightarrow {\bf r}_{j}, {\bf r}_{k} \; .
\end{equation}
The orbital functions used to construct the ansatz in Eq.~(\ref{ansatz_define}) are normalized Gaussians centered at the lattice sites and satisfy a linear Schr\"odinger equation with potential function of the form in Eq.~(\ref{paraboloid_pot}).

To break time-reversal symmetry the two sublattices $V_1({\bf r})$ and $V_2({\bf r})$ are rotated according to the  functions ${\bf h}_1(z), {\bf h}_2(z)$, respectively, i.e.
\begin{equation*}
\Delta n ({\bf r},z) = |\Delta n| \bigg[ 1 -  V_1({\bf r} - {\bf h}_1(z)) -  V_2({\bf r} - {\bf h}_2(z)) \bigg] \; .
\end{equation*}
The only restrictions we place upon the functions ${\bf h}_{1}(z), {\bf h}_{2}(z)$ is that they are smooth. 
{This is {quite} general.} {Since} {most researchers have studied periodic longitudinal variation,} {below} {we will assume ${\bf h}_1(z),{\bf h}_2(z)$ are periodic; 
but they need not be so.} In particular, we take
\begin{equation}
{\bf h}_{i}(z) = R_i \left( \cos\left( \Omega_i z + \chi_i \right) , \sin\left( \Omega_i z + \chi_i \right) \right) \; ,  ~ i = 1,2 
\end{equation}
where $R_i$ is the helix radius, $\Omega_i$ is the angular frequency of the oscillation, and $\chi_i$ a phase shift.

Next we move into a coordinate frame co-moving with with the $V_1({\bf r},z)$ sublattice by performing the change of variable
\begin{equation*}
{\bf r}' = {\bf r}  - {\bf h}_1(z) ~~~ , ~~~ z' = z \; ,
\end{equation*}
which (after the prime notation is dropped) gives the equation
\begin{equation}
\label{LSE_2}
i \frac{ \partial \psi}{\partial z} - i {\bf h}_1'(z) \cdot \nabla \psi+ \frac{1}{2 k_0} \nabla^2 \psi - \frac{k_0}{n_0}  \left( \Delta n({\bf r},z) + n_2 |\psi|^2 \right)\psi = 0  \; .
\end{equation}
Introducing the phase transformation
\begin{equation*}
\psi({\bf r},z) = \tilde{\psi}({\bf r},z) \exp\left(  \frac{ -i  \int_0^z | {\bf A}(\zeta)|^2 d \zeta }{2} \right) \; ,
\end{equation*}
for the pseudo-field defined by
$ {\bf A}(z) = - k_0 {\bf h}'_1(z) $
yields (after dropping the tilde notation)

\begin{equation}
\label{LSE_3}
i \frac{ \partial \psi}{\partial z} + \frac{1}{2 k_0} \left( \nabla + i  {\bf A}(z)\right)^2 \psi - \frac{k_0}{n_0}  \left( \Delta n({\bf r},z) +  n_2 |\psi|^2 \right)\psi = 0  \; .
\end{equation}
{We non-dimensionalize by
\begin{align*}
& x = l x' ~ , ~ y = l y' ~ , ~ z = z_* z' ~ , \\
 \sigma_x & =   l \sigma_x' ~, ~ \sigma_y = l \sigma_y' ~ , ~ \psi = \sqrt{I_*} \psi' ~ , 
\end{align*}
where $l$ is the distance between nearest-neighbor lattice sites, $z_* = 2 k_0 l^2$ is the characteristic propagation distance, $I_*$ is the peak input beam intensity, nonlinearity coefficient $\sigma=2 \gamma k_0l^2I_*$ and lattice {amplitude $V_0^2 =2k_0^2l^2 |\Delta n| /n_0$}. Dropping the $'$ notation, t}hese rescalings give the dimensionless equation
\begin{equation}
\label{LSE_4}
i \frac{ \partial \psi}{\partial z} + \left( \nabla + i {\bf A}(z)\right)^2 \psi - V({\bf r},z) \psi + \gamma |\psi|^2 \psi = 0  \; ,
\end{equation}
with 
the dimensionless potential 
\begin{equation*}
V({\bf r},z)  = V_0^2 \bigg[ 1 -  V_1({\bf r} ) -  V_2({\bf r} - \Delta {\bf h}_{21}(z)) \bigg] \; ,
\end{equation*}
where $\Delta {\bf h}_{21}(z) = {\bf h}_2(z) - {\bf h}_1(z).$

{In this paper we focus on periodic rotation in $z$.}
{The most general form of the dimensionless rotation functions is} {taken to be}
\begin{equation*}
 {\bf h}_i(z) = \eta_i \left( \cos\left( \frac{z}{\epsilon_i} + \chi_i \right) , \sin\left( \frac{z}{\epsilon_i} + \chi_i \right) \right) \; ,  ~ i = 1,2 
\end{equation*}
where $\eta_i \ge 0$ is the ratio of the helix radius to the distance between adjacent lattice sites, $\epsilon_i^{-1}$ is the helix rotation frequency, and $\chi_i$ is {a} phase shift.
The sign of $\gamma$ is taken to be positive corresponding to self-focusing Kerr nonlinear media \cite{kivshar_book} e.g. fused silica. 
To simplify Eq.~(\ref{LSE_4}) the phase transformation
$ \psi({\bf r},z) = \phi({\bf r},z) e^{ - i {\bf r} \cdot {\bf A}(z)}  $
is introduced and gives
\begin{equation}
\label{LSE_5}
i \frac{ \partial \phi}{\partial z} +  \nabla^2 \phi + {\bf r} \cdot {\bf A}_{z} \phi - V({\bf r},z) \phi + \gamma |\phi|^2 \phi = 0  \; .
\end{equation}
The dimensionless pseudo-field is given by {
\begin{equation}
\label{define_pseudo_field}
{\bf A}(z) = \kappa \left( \sin \left( \frac{z}{\epsilon_1}  + \chi_1 \right) , - \cos \left( \frac{z}{\epsilon_1}  + \chi_1 \right) \right) \; ,
\end{equation}
}where {$\kappa = k_0 \ell R_1 \Omega_1 =   \eta_1 / (2 \epsilon_1).$} 
{For each of the examples considered below we simply drop the subscript and call these parameters $ \eta, \epsilon$.}

To simplify the analysis of Eq.~(\ref{LSE_5}) a tight-binding approximation is applied. 
From a physical point of view this assumption is justified by the fact that many photonic experiments are performed in strong lattice regimes where $V_0^2 \gg 1$.
In the deep lattice limit the scalar field $\phi({\bf r},z)$ is well-approximated by {coupled} evanescent modes centered at the lattice sites. {Rigorous studies of these types of approximations were carried out in \cite{MCZ12}.~}

The translational symmetry of the lattice motivates the following ansatz \cite{AbZh10}
\begin{equation}
\label{ansatz_define}
\phi({\bf r},z) \sim \sum_v \left[ a_{\bf v}(z) \phi_{1,{\bf v}}({\bf r}) + b_{\bf v}(z) \phi_{2,{\bf v}}({\bf r},z)  \right] e^{i {\bf k} \cdot {\bf v} - i E z} \; ,
\end{equation}
{where the Gaussian functions $ \phi_{1,{\bf v}}$ and $\phi_{2,{\bf v}}$ satisfy the equations
\begin{align}
& \left( - \nabla^2   + V_{\rm loc}({\bf r} -{\bf v}) \right) \phi_{1,{\bf v}} = E  \phi_{1,{\bf v}}  \; , \\
& \left( - \nabla^2   + V_{\rm loc}({\bf r} -({\bf d} + {\bf v})-\Delta {\bf h}_{21}) \right) \phi_{2,{\bf v}} = E  \phi_{2,{\bf v}}  \; ,
\end{align}
for the local paraboloid potential
\begin{equation}
V_{\rm loc}({\bf r} )= V_0^2 \left( \frac{x^2}{\sigma_x^2} + \frac{y^2}{\sigma_y^2} \right) \; ,
\end{equation}
(see Appendix \ref{TBA_details} for more details).}

The two non-simple lattices configurations we consider in detail are honeycomb (see Fig.~\ref{honeycomb_fig}) and staggered square (see Fig.~\ref{center_sq_lattice}). The perfect {interpenetrating} square lattice, as shown in Fig.~\ref{center_sq_lattice}, is actually simple, but 
{{becomes} 
non-simple when there are necessarily different sublattices; e.g. by introducing a phase offset in one of the sublattices. Other lattice configurations can be considered with the method we present.}
The lattice sites for the honeycomb lattice are related via the characteristic vectors
\begin{equation}
\label{honey_lattice_vec_define}
{\bf v}_1 = \left( \frac{3}{2} ,  \frac{\sqrt{3}}{2} \right) , ~~  {\bf v}_2 = \left( \frac{3}{2} , -\frac{\sqrt{3}}{2} \right)  , ~~  {\bf d} =  \left( 1 , 0 \right)  \; ,
\end{equation}
where $|{\bf d}| = 1 $ is the distance between nearest neighbors.
The staggered square lattice is defined by the basis vectors
\begin{equation}
\label{square_lattice_vec_define}
{\bf v}_1 = \frac{1}{\sqrt{2}} \left( 1 ,  1 \right) , ~~  {\bf v}_2 = \frac{1}{\sqrt{2}} \left( 1 , -1 \right) \; ,
\end{equation}
and has a unit distance between neighboring lattice sites as well.

We substitute ansatz (\ref{ansatz_define}) into governing equation (\ref{LSE_5}) and multiply the resulting equation by {$\phi^*_{i,p}({\bf r},z), i = 1,2$}, where $^*$ indicates complex conjugation. The resulting equation is integrated over the entire domain. For a static isotropic lattice the interaction strengths are proportional to $\exp\left[ - (V_0 |{\bf v}|^2)/(4 \sigma) \right]$, where $|{\bf v}|$ is the distance between the different lattices sites. For a typical set of honeycomb lattice parameters (e.g. $V_0^2 = 50, \sigma = 1/2$) the nearest-neighbor coefficients are on the order of $\mathcal{O}(10^{-2}),$ while next-nearest neighbor interactions are considerably smaller at $\mathcal{O}(10^{-5}).$ In the case of the staggered squared lattice the next-nearest neighbor interaction is on the order of $\mathcal{O}(10^{-4}).$ For this reason we only consider self and nearest neighbor interactions and neglect all others.

\begin{figure}
\vspace{3mm}
\begin{tikzpicture}[node distance=1cm]

\begin{scope}[xshift=-3cm,yshift=1.3cm]
\node[regular polygon, regular polygon sides=6, shape aspect=0.5, minimum width=3cm, minimum height=1cm, draw,scale=1] (reg) {}; 
 \end{scope}

 \begin{scope}[xshift=-3cm,yshift=-1.3cm]
\node[regular polygon, regular polygon sides=6, shape aspect=0.5, minimum width=3cm, minimum height=1cm, draw,scale=1] (reg) {}; 
 \end{scope}
 
 \begin{scope}[xshift=-0.75cm,yshift=0cm]
\node[regular polygon, regular polygon sides=6, shape aspect=0.5, minimum width=3cm, minimum height=1cm, draw,scale=1] (reg) {}; 
 \end{scope}
 


 \draw (0,1.3) -- (0.75,2.6);
 \draw (0,-1.3) -- (0.75,-2.6);
 \draw (0.75,2.6) -- (2.25,2.6);
 \draw (0.75,0) -- (2.25,0);
 \draw (0.75,-2.6) -- (2.25,-2.6);
 \draw (-3.75,2.6) -- (-4.5,3.9);
 \draw (-3.75,-2.6) -- (-4.5,-3.9);  
 \draw (-2.25,2.6) -- (-1.5,3.9);
 \draw (-2.25,-2.6) -- (-1.5,-3.9); 
 \draw (0.75,2.6) -- (0,3.9);
 \draw (0.75,-2.6) -- (0,-3.9); 
    
 \fill (-1.5,1.3)  circle (2pt);
 \fill (-3.75,0)  circle (2pt);
 \fill (-1.5,-1.3)  circle (2pt);
 \fill (-3.75,2.6)  circle (2pt);
 \fill (-3.75,-2.6)  circle (2pt);

 \fill (0.75,0) circle (2pt);
 \fill (0.75,2.6) circle (2pt);
 \fill (0.75,-2.6) circle (2pt);

 \filldraw[fill=white] (-2.25,0) circle (2pt);
 \filldraw[fill=white] (-4.5,1.3) circle (2pt);
 \filldraw[fill=white] (-4.5,-1.3) circle (2pt);
 \filldraw[fill=white] (-2.25,2.6) circle (2pt);
 \filldraw[fill=white] (-2.25,-2.6) circle (2pt);
 
 \filldraw[fill=white]  (0,1.3)  circle (2pt);
 \filldraw[fill=white]  (0,-1.3)  circle (2pt);
 
 \draw [line width=0.3mm,->] (-3.755,0) -- (-2.4,0);
 \draw [line width=0.3mm,->] (-3.75,0) -- (-1.6,1.25);
 \draw [line width=0.3mm,->] (-3.75,0) -- (-1.6,-1.25);
 
 \draw (-3,-0.75) node () { $v_2$};
 \draw (-3,0.75) node () { $v_1$};
 \draw (-2.6,-0.2) node () { $d$};
 
 \draw (-4.2,0) node () { $a_{\rm m,1}$};
 \draw (-4.4,2.6) node () { $a_{\rm m+2,1}$};
 \draw (-4.4,-2.6) node () { $a_{\rm m-2,1}$};

 \draw (0.2,0) node () { $a_{\rm m,3}$};
 \draw (0.1,2.6) node () { $a_{\rm m+2,3}$};
 \draw (0.1,-2.6) node () { $a_{\rm m-2,3}$};
 
 \draw (-1.7,0) node () { $b_{\rm m,1}$};
 \draw (-1.6,2.6) node () { $b_{\rm m+2,1}$};
 \draw (-1.6,-2.6) node () { $b_{\rm m-2,1}$};
 
 \draw (-2.3,1.3) node () { $a_{\rm m+1,2}$};
 \draw (-2.3,-1.3) node () { $a_{\rm m-1,2}$};
 
 \draw (-3.8,1.3) node () { $b_{\rm m+1,0}$};
 \draw (-3.8,-1.3) node () { $b_{\rm m-1,0}$};
 \draw (0.7,1.3) node () { $b_{\rm m+1,2}$};
 \draw (0.7,-1.3) node () { $b_{\rm m-1,2}$};

\end{tikzpicture}

 \caption{{The honeycomb} lattice consists of two triangular sublattices $V_1({\bf r})$ and $V_2({\bf r})$ with minima (zeros) located at the white circles and black circles, respectively. The defining lattice vectors are given in Eq.~(\ref{honey_lattice_vec_define}).   \label{honeycomb_fig}}

\end{figure}
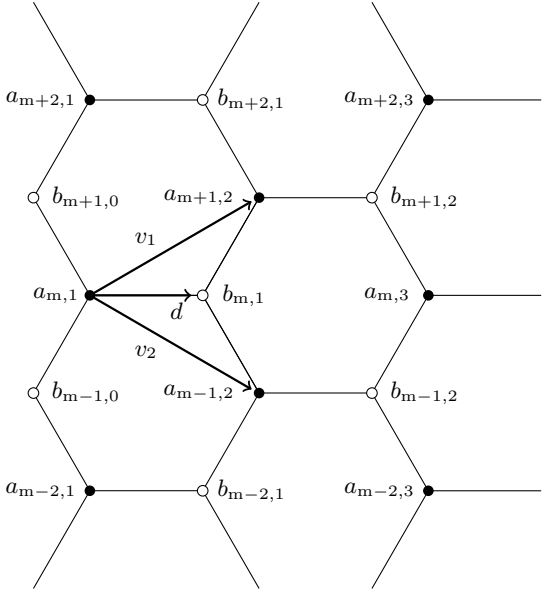

\subsection{Honeycomb Tight-binding Approximation}
\label{honey_TBA}

For the honeycomb lattice displayed in Fig.~\ref{honeycomb_fig} we derive a tight-binding approximation that takes into account the interactions of a Gaussian orbital with itself and its three nearest neighbors. In order to present the simplest picture possible, only the dominant terms are retained. The tight-binding approximation for this system is given by
\begin{align}
\nonumber & Honeycomb: \\ 
\label{honey_TBA_eq1}
& i \frac{d a_{mn}}{dz} + e^{i {\bf d} \cdot {\bf A}(z) + i \varphi(z)} \left( \mathcal{L}_-(z) b \right)_{mn} + \sigma |a_{mn}|^2 a_{mn} = 0 \; , \\
\label{honey_TBA_eq2}
& i \frac{d b_{mn}}{dz} + e^{-i {\bf d} \cdot {\bf A}(z) - i \varphi(z)} \left( \mathcal{L}_+(z) a \right)_{mn} + \sigma |b_{mn}|^2 b_{mn} = 0 \; ,
\end{align}
where $\varphi(z) = \int_0^z \left[ \Delta {\bf h}_{21}(\zeta) \cdot  {\bf A}_{\zeta}(\zeta) \right]d \zeta,$ and
\begin{align*}
 \left( \mathcal{L}_{-}(z) b \right)_{mn} =  & c_0(z) \left[  \mathbb{L}_0(z)  - i \mathbb{R}_0(z) \right]b_{mn} \\ \nonumber
+& c_1(z) \left[ \mathbb{L}_1(z) - i \mathbb{R}_1(z) \right]b_{m-1,n-1} e^{- i \theta_1(z)} \\ \nonumber
+& c_2(z) \left[  \mathbb{L}_2(z) - i \mathbb{R}_2(z)  \right]b_{m+1,n-1} e^{- i \theta_2(z)} \; ,
\end{align*}
\begin{align*}
 \left( \mathcal{L}_{+}(z) a \right)_{mn} =&  c_0(z) \mathbb{L}_0(z) a_{mn} \\ \nonumber
+& c_1(z) \mathbb{L}_1(z) a_{m+1,n+1} e^{ i \theta_1(z)} \\ \nonumber
+& c_2(z) \mathbb{L}_2(z) a_{m-1,n+1} e^{ i \theta_2(z)} \; ,
\end{align*}
{with} $ \theta_j(z) = ({\bf k} + {\bf A}(z)) \cdot {\bf v}_j$. Without loss of generality, we take ${\bf k} = 0$ \cite{AbCuMa2014}. The definitions of the coefficients  {$\mathbb{L}_j(z), \mathbb{R}_j(z), j=1,2,3$} are given in Appendix \ref{TBA_coefficients_honey}. The $z$-dependent coupling coefficient, $c_j(z)$, is a function that decays exponentially with {$V_0$}. The linear terms that are induced from the lattice potential and rotation effects are included in $\mathbb{L}_j(z)$. For a deep lattice ($V_0 \gg 1$), the magnitude of $c_j$ is small and the magnitude of $\mathbb{L}_j$ large, however {for the values of $V_0$ considered here,} their product is {numerically} $\mathcal{O}(1).$
When the two sublattices have the same rotation, i.e. $\Delta {\bf h}_{21} = 0$, this system reduces {to that}  considered in \cite{AbCuMa2014} by rescaling $Z = c_0 \mathbb{L}_0 z$, where all $c_j$ and $\mathbb{L}_j$ are now constant and $\mathbb{R}_j$ is zero {(see Appendix \ref{TBA_details}).}

Here we focus on the evolution of edge modes along the zig-zag edge of a semi-infinite strip domain, whose left side is shown in Fig.~\ref{honeycomb_fig}. As the beam evolves down the waveguide it is assumed to be well-confined inside the lattice region and negligibly small outside. The boundary conditions chosen to model edge modes on the left zig-zag boundary are: 
\begin{align*}
\nonumber
& a_{mn} = 0 ~~ {\rm for} ~~ n < 1 \; , ~~~~ b_{mn} = 0 ~~ {\rm for} ~~ n < 0 \; , \\
& a_{mn} \rightarrow 0 ~~ {\rm as} ~~ n \rightarrow \infty \; , ~~ b_{mn} \rightarrow 0 ~~ {\rm as} ~~ n \rightarrow \infty \; .
\end{align*}
The boundary conditions for edge modes on the right zig-zag edge mirror these.
We consider solutions of the form
\begin{equation}
\label{reduce_soln_dim}
 a_{mn}(z) = a_n(z; \omega) e^{i m \omega }  \; , ~~ 
 b_{mn}(z) = b_n(z; \omega) e^{i m \omega }  \; ,
\end{equation}
which reduce Eqs.~(\ref{honey_TBA_eq1})-(\ref{honey_TBA_eq2}) to 
\begin{align} 
 \label{1d_honey_TBA_eq1}
& i \frac{d a_{n}}{dz} + e^{i {\bf d} \cdot {\bf A}(z) + i \varphi(z)} \left( \widehat{\mathcal{L}}_-(z) b \right)_{n} + \sigma |a_{n}|^2 a_{n} = 0 \; , \\
\label{1d_honey_TBA_eq2}
& i \frac{d b_{n}}{dz} + e^{-i {\bf d} \cdot {\bf A}(z) - i \varphi(z)} \left( \widehat{\mathcal{L}}_+(z) a \right)_{n} + \sigma |b_{n}|^2 b_{n} = 0 \; ,
\end{align}
such that
\begin{align*}
 \left( \widehat{\mathcal{L}}_{-}(z) b \right)_{n} =&  c_0(z) \left[ \mathbb{L}_0(z)- i \mathbb{R}_0(z) \right] b_{n} \\ \nonumber
+& c_1(z) \left[ \mathbb{L}_1(z) - i \mathbb{R}_1(z) \right] b_{n-1} e^{- i \omega - i \theta_1(z)} \\ \nonumber
+& c_2(z) \left[ \mathbb{L}_2(z) - i \mathbb{R}_2(z) \right] b_{n-1} e^{i \omega - i \theta_2(z)} \; ,
\end{align*}
\begin{align*}
\left( \widehat{\mathcal{L}}_{+}(z) a \right)_{n} =&  c_0(z) \mathbb{L}_0(z) a_{n} \\ \nonumber
+& c_1(z) \mathbb{L}_1(z) a_{n+1} e^{i\omega +  i \theta_1(z)} \\ \nonumber
+& c_2(z) \mathbb{L}_2(z) a_{n+1} e^{- i \omega + i \theta_2(z)} \; .
\end{align*}

\subsection{Staggered square Tight-binding Approximation}
\label{sq_TBA}

Next we give a tight-binding approximation describing beam propagation in the staggered square lattice shown in Fig.~\ref{center_sq_lattice}. Only the {dominant} terms resulting from interactions of a Gaussian orbital with itself and its four nearest neighbors are considered. We point out that the indexing given for the square lattice in Fig.~\ref{center_sq_lattice} is different than that of the honeycomb lattice. Here the white lattice sites {(i.e. the `b' sites)} are all located at even points, and black sites  {(i.e. the `a' sites)} are located at odd positions. This is due to the fact that the underlying lattice is {rectangular.} The coupled system describing this {is}
\begin{align}
\nonumber  Staggered&~Square:  \\
 \label{square_TBA_eq1}
 i \frac{d a_{2m+1,2n+1}}{dz} &+ e^{ i \varphi(z)} \left( \mathcal{L}_-(z) b \right)_{2m+1,2n+1} \\ \nonumber
& ~~~~~~~~~  + \sigma |a_{2m+1,2n+1}|^2 a_{2m+1,2n+1} = 0 \; , \\
 \label{square_TBA_eq2}
 i \frac{d b_{2m,2n}}{dz} & + e^{- i \varphi(z)} \left( \mathcal{L}_+(z) a \right)_{2m,2n} \\ \nonumber
& ~~~~~~~~~~~~~~~ + \sigma |b_{2m,2n}|^2 b_{2m,2n} = 0 \; ,
\end{align}
where $\varphi(z) = \int_0^z \left[ \Delta {\bf h}_{21}(\zeta) \cdot  {\bf A}_{\zeta}(\zeta) \right]d \zeta,$ and
\begin{align*}
& \left( \mathcal{L}_{-}(z) b  \right)_{2m+1,2n+1}  = \\ \nonumber
 &~~~~ c_1(z) \left[  \mathbb{L}_1(z) - i \mathbb{R}_1(z) \right] b_{2m+2,2n+2} e^{ i \theta_1(z)} \\ \nonumber
& + c_{-1}(z) \left[ \mathbb{L}_{-1}(z) - i \mathbb{R}_{-1}(z) \right] b_{2m,2n} e^{- i \theta_1(z)} \\ \nonumber
&+ c_2(z) \left[ \mathbb{L}_2(z) - i \mathbb{R}_2(z)\right] b_{2m,2n+2} e^{ i \theta_2(z)} \\ \nonumber
&+ c_{-2}(z) \left[ \mathbb{L}_{-2}(z) - i \mathbb{R}_{-2}(z) \right] b_{2m+2,2n} e^{- i \theta_2(z)} \; ,
\end{align*}
\begin{align*}
 \left( \mathcal{L}_{+}(z) a \right)_{2m,2n} =&  c_{-1}(z) \mathbb{L}_{-1}(z) a_{2m+1,2n+1}  e^{i \theta_1(z)}  \\ \nonumber
+& c_1(z) \mathbb{L}_1(z) a_{2m-1,2n-1} e^{- i \theta_1(z)} \\ \nonumber
+& c_{-2}(z) \mathbb{L}_{-2}(z) a_{2m-1,2n+1} e^{ i \theta_{2}(z)} \\ \nonumber 
+& c_2(z) \mathbb{L}_{2}(z) a_{2m+1,2n-1} e^{- i \theta_{2}(z)}\; ,
\end{align*}
with $ \theta_j(z) = ({\bf k} + {\bf A}(z)) \cdot {\bf v}_j$. 
The definitions for the coefficients  {$\mathbb{L}_j(z), \mathbb{R}_j(z), j=\pm1,\pm2 $}  are given in Appendix \ref{TBA_coefficients_square}. Again, for simplicity we take ${\bf k} = {\bf 0}.$

We focus on the edge modes propagating along the edge of a semi-infinite strip domain. For localized modes traveling on the left side of the domain we take the boundary conditions
\begin{align*}
\nonumber
& a_{2m+1,2n+1} = 0 ~~ {\rm for} ~~ n < 0 \; , ~~~~ b_{2m,2n} = 0 ~~ {\rm for} ~~ n < 0 \; , \\
& a_{2m+1,2n+1} \rightarrow 0 ~~ {\rm as} ~~ n \rightarrow \infty \; , ~~ b_{2m,2n} \rightarrow 0 ~~ {\rm as} ~~ n \rightarrow \infty \; .
\end{align*}
The boundary conditions on the right side mirror these. 
We take solutions of the form
\begin{align}
\nonumber
& a_{2m+1,2n+1}(z) = a_{2n+1}(z; \omega) e^{i (2m+1) \omega } \; , \\
& b_{2m,2n}(z) = b_{2n}(z; \omega) e^{i 2m \omega } \; ,
\end{align}
for real $\omega$ which yield the following coupled system 
\begin{align}
\label{1d_square_TBA_eq1}
& i \frac{d a_{2n+1}}{dz} + e^{ i \varphi(z)} \left( \widehat{\mathcal{L}}_-(z) b \right)_{2n+1} + \sigma |a_{2n+1}|^2 a_{2n+1} = 0 \; , \\
\label{1d_square_TBA_eq2}
& i \frac{d b_{2n}}{dz}  + e^{- i \varphi(z)} \left( \widehat{\mathcal{L}}_+(z) a \right)_{2n} + \sigma |b_{2n}|^2 b_{2n} = 0 \; .
\end{align}
where
\begin{align*}
\nonumber
 \left( \widehat{\mathcal{L}}_{-}(z) b \right)_{2n+1} =& 
   c_1(z) \left[  \mathbb{L}_1(z) - i \mathbb{R}_1(z) \right]   b_{2n+2} e^{i \omega + i \theta_1(z)} \\ \nonumber
+& c_{-1}(z) \left[ \mathbb{L}_{-1}(z) - i \mathbb{R}_{-1}(z) \right] b_{2n} e^{- i \omega - i \theta_1(z)} \\ \nonumber
+& c_2(z) \left[ \mathbb{L}_2(z) - i \mathbb{R}_2(z)\right]b_{2n+2} e^{- i \omega +  i \theta_2(z)} \\ \label{1d_square_linear_terms1}
+& c_{-2}(z) \left[ \mathbb{L}_{-2}(z) - i \mathbb{R}_{-2}(z) \right] b_{2n} e^{i \omega - i \theta_2(z)} \; , 
\end{align*}
\begin{align*}
 \left( \widehat{\mathcal{L}}_{+}(z) a \right)_{2n} =&  c_{-1}(z) \mathbb{L}_{-1}(z) a_{2n+1}  e^{i \omega + i \theta_1(z)}  \\ \nonumber
+& c_1(z) \mathbb{L}_1(z) a_{2n-1} e^{- i \omega - i \theta_1(z)} \\ \nonumber
+& c_{-2}(z) \mathbb{L}_{-2}(z) a_{2n+1} e^{- i \omega +  i \theta_{2}(z)} \\ \nonumber 
+& c_2(z) \mathbb{L}_{2}(z) a_{2n-1} e^{ i \omega - i \theta_{2}(z)}\; .
\end{align*}

\section{Linear Floquet Bands and Edge State Dynamics}
\label{linear_theory}
In this section we consider a low amplitude linear limit (i.e. $|a_{mn}|^2, |b_{mn}|^2 \approx 0$) of the full nonlinear systems given in the previous section. The dispersion relation $\alpha(\omega)$ is computed numerically via Floquet theory \cite{Eastham1973}. The Floquet multipliers $\gamma = e^{- i \alpha T + i 2 \pi \tau}, \tau \in \mathbb{Z}$ are obtained from the eigenvalues  of the fundamental matrix solution at $z = T$, where $T$ is the period of the lattice. A fourth-order Runge-Kutta method is used to integrate. For all band diagrams below, 40 lattice sites {(in both $a_n$ and $b_n$)} are used. 
 The Floquet exponent is calculated up to an additive constant by
\begin{equation}
\label{floquet_exponent}
\alpha(\omega) =  \frac{i \log (\gamma(\omega))}{T} - \frac{2 \pi \tau}{T} \; , ~~ \tau \in \mathbb{Z} \; .
\end{equation}
{To be specific we will} focus on five different rotation patterns among the sublattices: 
\begin{itemize}
\item same rotation, same phase
\begin{equation}
\label{same_rot_func}
{\bf h}_{2}(z) = {\bf h}_{1}(z) = \eta \left( \cos\left( \frac{z}{\epsilon}  \right) , \sin\left( \frac{z}{\epsilon} \right) \right) \; ,
\end{equation}
\item different radii, same phase 
{\begin{equation}
\label{diff_rad_rot_func}
{\bf h}_{2}(z) = R_{a} {\bf h}_{1}(z) = R_a \eta \left( \cos\left( \frac{z}{\epsilon}  \right) , \sin\left( \frac{z}{\epsilon} \right) \right) \; , ~~ R_a < 1
\end{equation}}
\item $\pi$-phase offset rotation
\begin{equation}
\label{pi_phase_rot_func}
{\bf h}_{2}(z) = {\bf h}_{1}(z + \epsilon \pi) =  - \eta \left( \cos\left( \frac{z}{\epsilon}  \right) , \sin\left( \frac{z}{\epsilon}  \right) \right) \; ,
\end{equation}
\item counter rotation 
\begin{equation}
\label{counter_rot_func}
{\bf h}_{2}(z) = {\bf h}_{1}(-z) = \eta \left( \cos\left( \frac{z}{\epsilon}  \right) , -\sin\left( \frac{z}{\epsilon} \right) \right) \; ,
\end{equation}
\item different frequency, same phase
\begin{equation}
\label{diff_freq_rot_func}
{\bf h}_{2}(z) = {\bf h}_{1}(2z) = \eta \left( \cos\left( \frac{2 z}{\epsilon}  \right) , \sin\left( \frac{2 z}{\epsilon} \right) \right) \; .
\end{equation}
\end{itemize}
We point out that only ${\bf h}_2(z)$ is adjusted to take into account the non-synchronized sublattice motion above.
{The pseudo-field, ${\bf A}(z),$ is defined in Eq.~(\ref{define_pseudo_field}).}
The physical parameters chosen {in the simulations below} are presented in Table~\ref{table1} and reflect the experimental setup used in \cite{RechtsSegev2013}. For these values the dimensionless parameters are $\epsilon \approx 0.75/\pi$ and $V_0^2 \approx 45.$
Moreover, one unit in the dimensionless $z$ is equal to 6.5 mm in physical units.
\begin{table}
\centering
  \begin{tabular}{ | c | c | c | c | c | c |}
    \hline
      $\ell~(\mu {\rm m})$ & $n_0$ & $\lambda~({\rm nm})$ & $\Omega~({\rm rad/cm})$ & $|\Delta n|$  \\ \hline
      15 & 1.45 & 633 & 2$\pi$ & 7$\times 10^{-4}$  \\ \hline
  \end{tabular}
\caption{ \label{table1}Physical parameters.}
\end{table}

\subsection{Honeycomb Floquet bands}

\begin{figure} [ht]
\centering
\includegraphics[scale=.75]{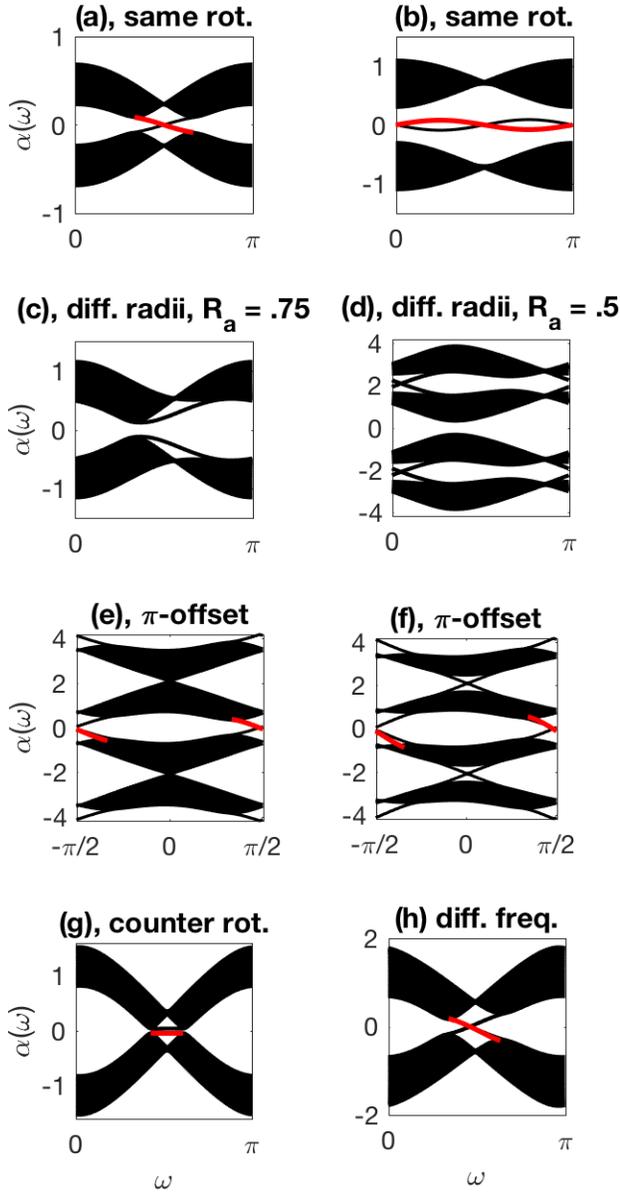}
\caption{(Color online) The honeycomb lattice linear band structure (\ref{floquet_exponent}) of Eqs.~(\ref{1d_honey_TBA_eq1})-(\ref{1d_honey_TBA_eq2}) {for different sublattice rotation patterns. The value of  $\eta$ is: (a-d) $2/3$, (e) $1.7/15$, (f) $2/15$, (g) 1/10, and (h) $2/15$. The lattice parameters are $V_0^2 = 45, \epsilon = 0.75/\pi, {\rm and} ~\sigma_x = \sigma_y = 0.3$,} except (b) where $\sigma_x = 0.5, \sigma_y = 0.25$. Red curves correspond to the asymptotic solution given in Eq.~(\ref{asym_floq_exp}). }
\label{linear_honey_bands}
\end{figure}

In this section we explore the linear band structure induced by various rotation patterns for the semi-infinite honeycomb lattice shown in Fig.~\ref{honeycomb_fig}. The Floquet exponents defined in (\ref{floquet_exponent}) are determined for the {linear} one-dimensional system given in Eqs.~(\ref{1d_honey_TBA_eq1})-(\ref{1d_honey_TBA_eq2}).
The first Brillouin zone for several Floquet bands {is} shown in Fig.~\ref{linear_honey_bands}.
Each band structure consists of bulk/extended (solid regions) and edge/localized (curves) modes. {The parameters used in the simulations are given in Fig. ~\ref{linear_honey_bands}.}

\begin{figure} [h]
\includegraphics[scale=.52]{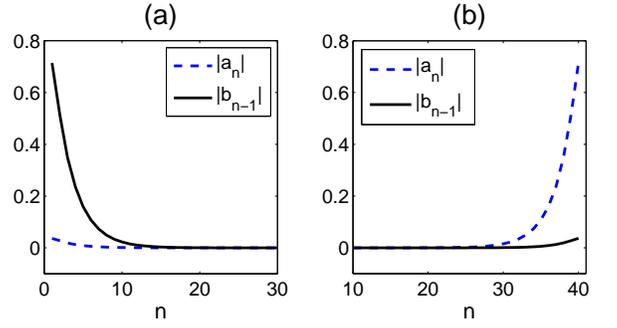}
\caption{Honeycomb lattice edge modes corresponding to the (a) left  and (b) right zig-zag edges of Fig.~\ref{honeycomb_fig}. The lattice parameters are the same as those used to generate Fig.~\ref{linear_honey_bands}(a) at $\omega = 5 \pi /8$ and correspond to the curves with (a) negative and (b) positive slope.}
\label{plot_eigenmodes}
\end{figure}

The first case we consider is that of isotropic waveguides rotating in phase with each other {(\ref{same_rot_func})} [see Fig.~\ref{linear_honey_bands}(a)]. 
{We find this case supports} unidirectional edge states. Typical eigenfunctions from both curve branches are displayed in Fig.~\ref{plot_eigenmodes}. The eigenmode shown in Fig.~\ref{plot_eigenmodes}(a) corresponds to the left zig-zag edge and has strictly negative group velocity i.e. $\alpha'(\omega) < 0 $. 
The situation is reversed for edge modes on the right zig-zag edge, shown in Fig.~\ref{plot_eigenmodes}(b), that have positive group velocity ($\alpha'(\omega) > 0$). Taken together, this indicates that the edges modes on a finite domain propagate along the boundary of the lattice in a counter-clockwise fashion, the same direction of the rotating waveguides. Moreover, the outer boundary modes [$|b_0|$ in Fig.~\ref{plot_eigenmodes}(a)] are found to have approximately 20 times larger magnitude than the inner boundary mode [$|a_1|$ in Fig.~\ref{plot_eigenmodes}(a)].

An asymptotic theory, given in Appendix \ref{Asymptotic_Analysis}, yields (to leading order) left zig-zag edge modes 
\begin{equation}
\label{linear_mode_solns}
a_{mn}(z;\omega_0) = 0 \; , ~~ b_{mn}(z;\omega_0) = C(z) b_n^s(\omega_0) e^{i m \omega_0} \; ,
\end{equation}
evaluated at the wavemode $\omega_0,$ where
$ C(z) = C_0 \exp{ \left( -i \epsilon \tilde{\alpha} z \right)} $
for constant $C_0$ and $\tilde{\alpha}(\omega_0)$ {is found explicitly in terms of integrals; it is given} in Eq.~(\ref{asym_floq_exp}). The theory is valid in the regime where the waveguides are rapidly oscillating {($|\epsilon| \ll 1$)} and describes edge modes in the central gap {near $\alpha = 0$; we} refer to this gap as the {\it{Floquet center}}.
Where applicable, the asymptotic solutions are compared against those computed numerically.

Next we consider when the waveguides are {rotating in phase with each other {(\ref{same_rot_func})} but are} anisotropic ($\sigma_x \not= \sigma_y$). The band structure in Fig.~\ref{linear_honey_bands}(b) has lost its topologically protected edge {character} and we {expect} backscatter. This observation comes from the fact that the slope of edge curves are no longer sign-definite in one Brillouin zone. We point out that this structure only occurs when $\sigma_x > \sigma_y$ i.e. when the waveguides are elliptical with the major axis in the $x$-direction (also $n$-direction).
 In terms of the results presented in \cite{AbCuMa2014}, this regime corresponds to $\rho < 1/2.$ 

We now examine when the two sublattices have the same phase, but one of the waveguides has a radius that is {three-quarters or half the size} of the other {(\ref{diff_rad_rot_func})} {: Fig.~\ref{linear_honey_bands}(c-d)}.
{In the first case ($R_a = 3/4$), the crossing modes observed in Fig.~\ref{linear_honey_bands}(a) have deformed into the non-crossing curves seen in Fig~\ref{linear_honey_bands}(c), which still have {a sign-definite} slope.}
The fundamental Floquet exponent, $\tau = 0$, as well as the first non-fundamental Floquet bands, $\tau = \pm 1$, defined in Eq.~(\ref{floquet_exponent}) are shown in Fig.~\ref{linear_honey_bands}(d). The point at which these two branches meet is $\alpha = \pm \pi / T$. {We refer to this point as the {\it Floquet edge}}. {We see that there is a family of edge modes that live in this Floquet edge gap; and  the} modes in this gap are unidirectional.

\begin{figure} [h]
\includegraphics[scale=.38]{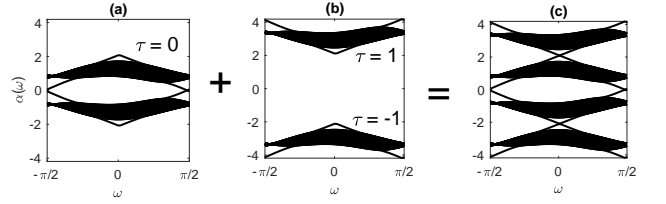}
\caption{Floquet bands (\ref{floquet_exponent}) used to create the $\pi$-phase offset band structure shown in Fig.~\ref{linear_honey_bands}(f). The fundamental ($\tau = 0$) and its periodic extensions ($\tau = \pm 1$) are combined at the Floquet edge $\alpha = \pm \pi/T \approx \pm2.09$.}
\label{floq_edge_combine}
\end{figure}

Now let us consider a $\pi-$phase offset between the two sublattices {(\ref{pi_phase_rot_func})}. The band structure for this arrangement is shown in Figs.~\ref{linear_honey_bands}(e-f). Included is the fundamental Floquet band ($\tau = 0$) as well as its periodic extensions, $\tau = \pm 1.$ Presentation of the different branches of the Floquet bands in Fig.~\ref{floq_edge_combine} illustrates how {one can understand the meeting/merging of the the inter-band curves}  near the Floquet edge. {At sufficiantly} {small helix radii the band structure consists of bulk bands separated by gaps and {exhibits} no edge modes.}
As the radius of the lattice helix, measured by $\eta$, increases the gap between the adjacent Floquet bands begins to close until the bands ``kiss'' each other at a single point in Fig.~\ref{linear_honey_bands}(e). This point marks a transition from which any additional driving introduces a new family of topologically protected edge modes at the Floquet edge [see Fig.~\ref{linear_honey_bands}(f)]. We point out that these edge states are distinct from those at the Floquet center. {Another feature that we observe in this case, unlike the case in Fig.~\ref{linear_honey_bands}(a), is that the edge states do not remain as $\epsilon \rightarrow 0$} {(This is shown in Fig.~\ref{finite_eps_evolve}).}

{The next case examined is that of counter rotation (\ref{counter_rot_func}). The bands corresponding to this case are shown in Fig.~\ref{linear_honey_bands}(g). We find two flat bands, separated by a small gap, that correspond to non-traveling states. The opposing motion of the two sublattices {evidently} cancel each other out yielding zero net movement{; i.e. essentially a stationary state}. 
The final scenario is when one sublattice has twice the frequency of the other sublattice {(\ref{diff_rad_rot_func})}. The band structure, displayed in Fig.~\ref{linear_honey_bands}(h), exhibits a set of topologically protected edge modes. We find that this configuration may possess a weak instability}{[, namely, $|{\rm Im}~ \alpha | = \mathcal{O}(10^{-3})$}; {i.e. it would be a weak exponentially growing/decaying mode} that would only manifest itself after a long {distance}. 


\begin{figure} [ht]
\includegraphics[scale=.78]{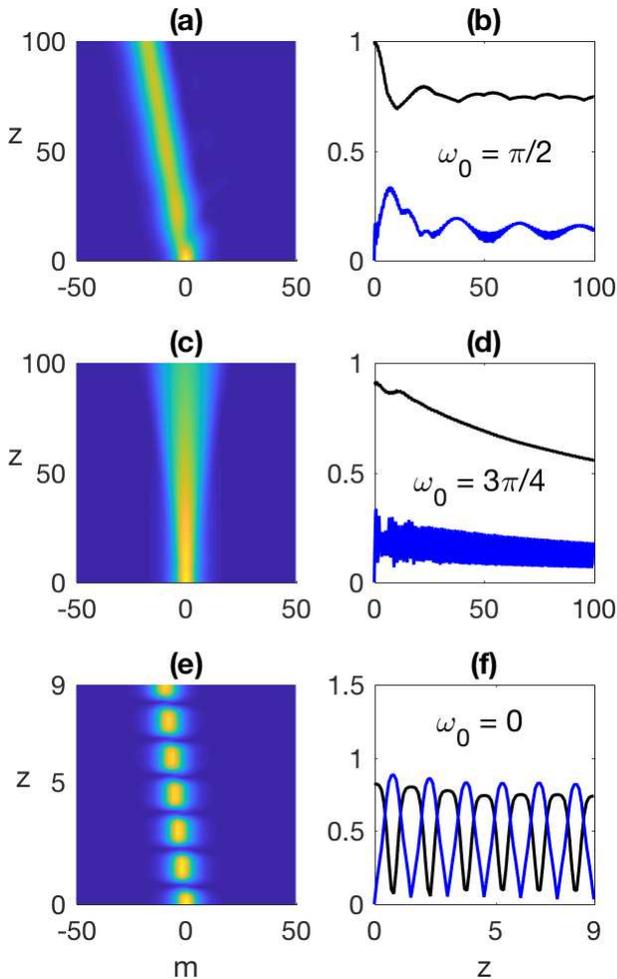}
\caption{(Color online) (Left column) Evolution of linear edge mode magnitude $\left| b_{m,0}(z) \right|$ in the honeycomb lattice. (Right column) Corresponding evolution of $\max_{m,n} |a_{mn}(z)|$ (blue) and $\max_{m,n} |b_{mn}(z)|$ (black). The corresponding Floquet bands are: (a-b) Fig.~\ref{linear_honey_bands}(a), (c-d) Fig.~\ref{linear_honey_bands}(b), and (e-f) Fig.~\ref{linear_honey_bands}(f).}
\label{linear_honeycomb_evolution}
\end{figure}

\subsection{Linear edge mode dynamics in honeycomb lattice}

In this section we explore the dynamics of the linear edge states found in Fig.~\ref{linear_honey_bands}.
Specifically, we numerically integrate the full tight-binding system (\ref{honey_TBA_eq1})-(\ref{honey_TBA_eq2}) {corresponding to} solutions of the form given in Eq.~(\ref{linear_mode_solns}). In all simulations we {omit nonlinearity:} $\sigma = 0.$
The solutions are initialized at a chosen frequency, $\omega_0,$ by
\begin{equation}
\label{linear_HC_IC}
a_{mn}(0) = 0 \; , ~~~~~~ b_{mn}(0) = {\rm sech}\left( \mu m \right) b_n(\omega_0) e^{i m \omega_0 } \; ,
\end{equation}
where $\mu = 0 .3$ {denotes the slowly varying envelope} and $b_n(\omega_0)$ is the numerically computed mode whose Floquet bands are shown in Fig.~\ref{linear_honey_bands}. For consistency, we normalize all edge modes so the two-norm is one, namely $<b_n,b_n> = 1$ via the discrete inner product $<f_n,g_n> = \sum_n f_n^* g_n.$
We take periodic boundary conditions in the $m$-direction and zero boundary conditions in the $n$-direction.
For all $z$-evolutions shown in this paper we take $N = 100$ sites in the $n$-direction to ensure the mode 
{has} sufficiently decayed.
The system is integrated using a fourth-order Runge-Kutta method.
 
Several typical evolutions are displayed in Fig.~\ref{linear_honeycomb_evolution}. In the left column we show the solution magnitude at the {left} boundary of {the} domain {i.e. $n = 0$}. In the right column the maximum magnitude for both $a_{mn}(z)$ {(in blue)} and $b_{mn}(z)$ {(in black)} over the entire domain are given. In Figs.~\ref{linear_honeycomb_evolution}(a-b) the propagation of a topologically protected edge mode, whose dispersion curves are given in Fig.~\ref{linear_honey_bands}(a), is shown. The traveling mode moves in the negative direction at a constant speed. At a point of inflection, $\alpha''(\omega) = 0$, a method of stationary phase calculation shows that the edge mode decays like $\sim z^{-1/3}$. Next in Figs.~\ref{linear_honeycomb_evolution}(c-d) a stationary edge state that is {\it not} topologically protected [corresponding to the bands in Fig.~\ref{linear_honey_bands}(b)] is shown. In contrast to the previous case, this state is diffracting and losing amplitude at a rapid pace. Asymptotically, when $\alpha''(\omega) \not=0$ the mode decays like $\sim z^{-1/2}$. 

The next mode we consider is located near the Floquet edge in Fig.~\ref{linear_honey_bands}(f) and corresponds to $\pi$-offset rotation. Unlike the previous two cases, whose energy is always in the $b_{m0}$ lattice sites, this solution [shown in Figs.~\ref{linear_honeycomb_evolution}(e-f)] is observed to oscillate back-and-forth between the $b_{m0}$ and $a_{m1}$ lattice sites. Energy is regularly transferred back-and-forth each cycle of the lattice waveguides (here the period is $T = 1.5$). A similar evolution pattern (not shown here) was found to occur for the edge state mode corresponding to different sublattice radii whose Floquet bands are shown in Fig.~\ref{linear_honey_bands}(d).

Floquet theory does tell us something about this coupled-mode dynamic. 
Typically, the 1D honeycomb Eqs.~(\ref{1d_honey_TBA_eq1})-(\ref{1d_honey_TBA_eq2}) are integrated over {the} period {$[0,T]$} to find the fundamental matrix solution. The corresponding edge mode eigenfunction [similar to that in Fig.~\ref{plot_eigenmodes}(a)] has considerably more energy in the outer $b_{n}$ sites than those of $a_{n}$. If we instead calculate the Floquet multipliers {over the interval $[T/2,3T/2]$}, then the energy is primarily concentrated in the $a_{n}$ sites, rather than $b_{n}$. From this observation we infer that these solutions are not of the form given in Eq.~(\ref{linear_mode_solns}), and instead resemble $a_{n}(z) = A(z)r^{n}$ and $b_{n}(z) = B(z)r^n $ for $|r| <1, n = 0,1,2, \dots$, where $A(z)$ and $B(z)$ are $T$-periodic envelopes.
In other words, the modes are truly coupled ($a_n \not= 0$) and can not be assumed to be scalar ($a_n \approx 0$).


\subsection{Staggered square Floquet bands}

\begin{figure} [h]
\includegraphics[scale=.74]{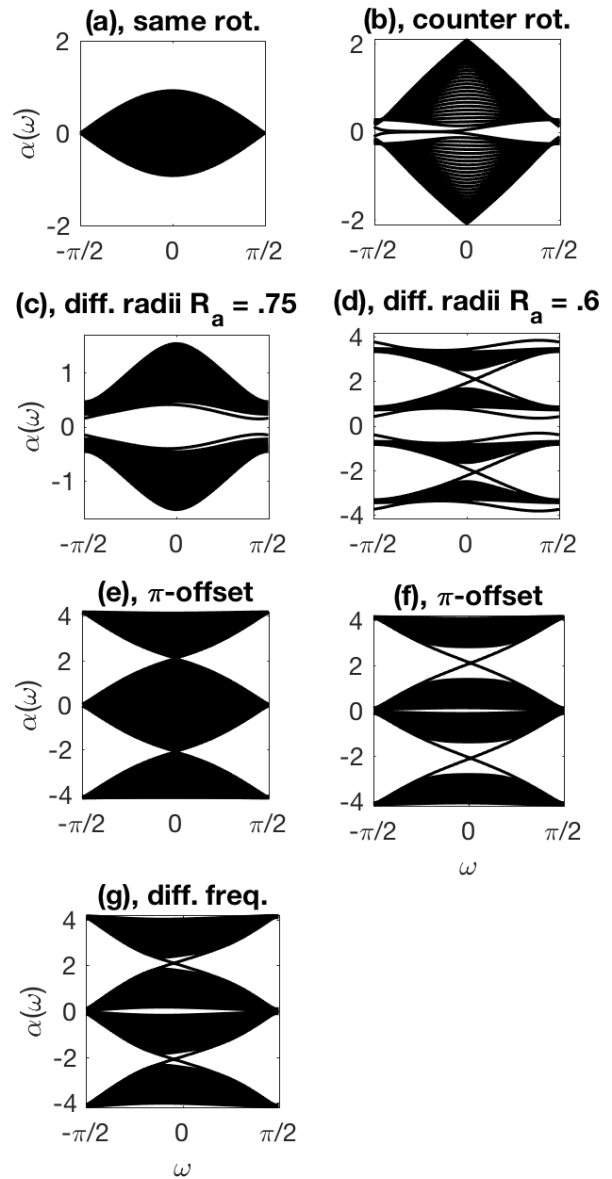}
\caption{The staggered square lattice linear band structure (\ref{floquet_exponent}) of Eqs.~(\ref{1d_square_TBA_eq1})-(\ref{1d_square_TBA_eq2}) for {different rotation patterns. The value of $\eta$ is: (a) $2/3$,  (b) $2/15$,  (c) $2/3$,  (d) $2/3$, (e) $1.1/15$, (f) $1.7/15$, (g) $2/15$. The lattice parameters are: $V_0^2 = 45, \epsilon = 0.75/\pi,$ and $ \sigma_x = \sigma_y = 0.3.$}}
\label{linear_square_bands}
\end{figure}

In this section the Floquet bands for the staggered square lattice in Fig.~\ref{center_sq_lattice} are computed for {the rotation patterns described earlier in  equations (\ref{same_rot_func})-(\ref{diff_freq_rot_func}).} The Floquet exponent (\ref{floquet_exponent}) is computed from 1D staggered square system (\ref{1d_square_TBA_eq1})-(\ref{1d_square_TBA_eq2}).
Several band structures are shown in Fig.~\ref{linear_square_bands}. 

The first lattice configuration we examine is when the two sublattices rotate in phase with each other {given in equation (\ref{same_rot_func}).} In this case, the system given in Eqs.~(\ref{1d_square_TBA_eq1})-(\ref{1d_square_TBA_eq2}) is degenerate and reduces {to} a single equation. In terms of the lattice structure, the lattice sites form a simple lattice. There are no edge modes at all{, as indicated in} Fig.~\ref{linear_square_bands}(a){; there are} only extended bulk modes. 
One way to create a non-simple lattice configuration is to make the radius of one sublattice smaller than the other.
The dispersion bands for this scenario are shown in Figs.~\ref{linear_square_bands}(c) and \ref{linear_square_bands}(d) where the radii of one sublattice is $75\%$ and $60\%$, respectively, the size of the other.
As the radius disparity grows, localized edge state curves manifest themselves in the gap at the Floquet edge. {This latter band structure is distinguished from all other cases in that  for certain positive $\omega$ values there exist two different Floquet exponents: one topologically-protected near the Floquet edge and the other non-protected near the Floquet center. In most} {of the topological systems we investigated,} { either one or the other mode types exist at a fixed $\omega$, but not both.}

The two rotation patterns we consider next are counter rotation {(see equation (\ref{counter_rot_func})} and $\pi$-phase offset rotation {(see equation (\ref{pi_phase_rot_func}).}
First, we examine the counter rotation case whose Floquet bands are shown in Fig.~\ref{linear_square_bands}(b). If the radius of the sublattices are driven hard enough, then, similar to the honeycomb case, the bulk band splits apart and inside the central gap 
 {an} 
edge mode corresponding to a nearly flat band is found to exist. Next we consider a $\pi$-offset rotation. We point out that this scenario {was} 
explored in \cite{Leykam2016, Leykam2016A}. Qualitatively, our results agree with theirs, namely there is a transition point in the topological structure of the bands that is displayed in Fig.~\ref{linear_square_bands}(e). Below this transition point, in a weak rotation regime, the band structure takes {a} ``trivial'' form similar to that seen in Fig.~\ref{linear_square_bands}(a). If, on the other hand, the spiral radius is increased beyond this threshold a set of topologically protected edge modes {emerge} and give the ``nontrivial'' band structure shown in Fig.~\ref{linear_square_bands}(f). This transition point in band structure resembles the honeycomb cases given in Figs.~\ref{linear_honey_bands}(e-f).

The final case we consider is that of different frequency between the two sublattices {(see equation (\ref{diff_freq_rot_func}).} {For a large enough helix radius, edge states are found to occur near the Floquet edge. Similar to the honeycomb case in Fig.~\ref{linear_honey_bands}(h), these states appear to possess a weak instability [$|{\rm Im}~\alpha| = \mathcal{O}(10^{-3})$] that could become relevant over very long propagation distances.}


At this point we summarize the commonalities in the linear band structures between the honeycomb and staggered square lattices. In the case of same rotation patterns we find that edge modes are present if the underlying lattice is {fundamentally} non-simple (like honeycomb, {or staggered with different sublattice size or frequency}). Taking sufficiently different radii among the sublattices creates edge states at the Floquet edges, for both lattice types.
Stationary (or nearly stationary) edge modes are generated in the case of counter-rotating sublattices.
A family of edge modes is found to exist between different branches of Floquet bands (\ref{floquet_exponent}) for $\pi$-offset rotation when the {waveguide parameters exceed} 
a certain threshold. When the two sublattices have different frequencies 
edge modes can be {found} 
{(with a possible weak instability)}.

\subsection{Linear edge mode dynamics in staggered square lattice}

\begin{figure} [ht]
\includegraphics[scale=.65]{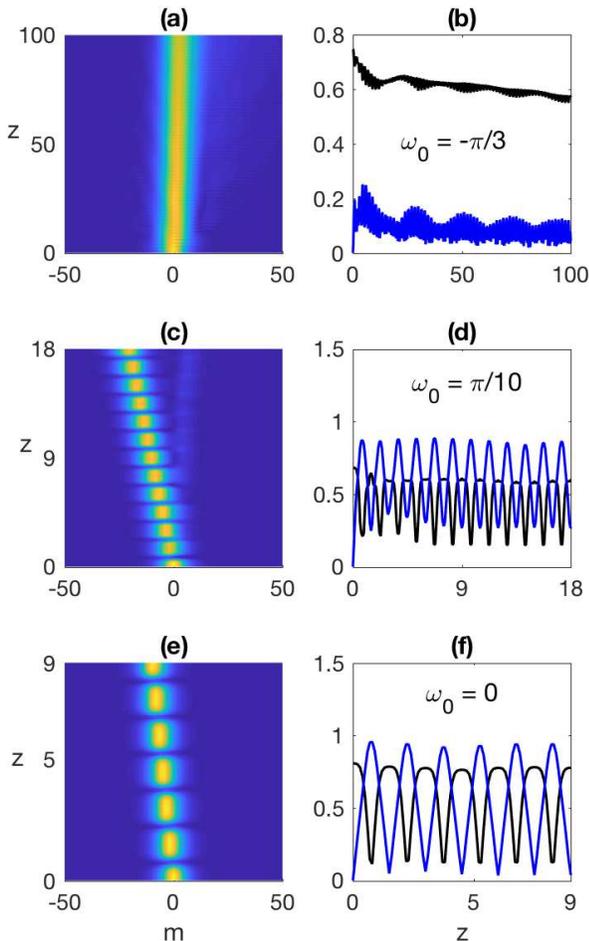}
\caption{(Color online) (Left column) Evolution of linear edge mode magnitude $\left| b_{m,0}(z) \right|$ in the staggered square lattice. (Right column) Corresponding evolution of $\max_{m,n} |a_{mn}(z)|$ (blue) and $\max_{m,n} |b_{mn}(z)|$ (black). The corresponding Floquet bands are: (a-b) Fig.~\ref{linear_square_bands}(b), (c-d) Fig.~\ref{linear_square_bands}(d), and (e-f) Fig.~\ref{linear_square_bands}(f).}
\label{linear_square_evolve}
\end{figure}

In this section we propagate the linear modes associated with the dispersion curves shown in Fig.~\ref{linear_square_bands}. 
The governing equations (\ref{square_TBA_eq1})-(\ref{square_TBA_eq2}) are integrated using initial conditions of the form
\begin{equation}
\label{linear_Sq_IC}
a_{2m+1,2n+1}(0) = 0 \; , ~~ b_{2m,2n}(0) = {\rm sech}\left(  2m \mu \right) b_{2n}(\omega_0) e^{i 2 m \omega_0 } \; ,
\end{equation}
where $\mu = 0 .3$ and the decaying mode $b_{2n}$ is numerically computed from the 1D system (\ref{1d_square_TBA_eq1})-(\ref{1d_square_TBA_eq2}).
As was the case in honeycomb lattice, we take $N = 100$ lattice sites in the $n$-direction for both $ b_{2m,2n}$ and $a_{2m+1,2n+1}$. 
For all cases considered here we display the edge magnitude as well as the maximum magnitude over the domain {as a function of $z$}.

The first nontrivial case we consider is that of counter-rotation whose corresponding dispersion curves are shown in Fig.~\ref{linear_square_bands}(b). The Floquet bands are nearly flat and so very little translation of the localized mode is {expected}.
~The edge mode evolution is shown in Figs.~\ref{linear_square_evolve}(a-b) and indeed the mode is well-localized{, has a small positive velocity and most of the energy is in the `$b$' mode.} Next we look at the case of different radii among the sublattices corresponding to the bands  displayed in Fig.~\ref{linear_square_bands}(d). The edge wave propagation and magnitude evolution is shown in Figs.~\ref{linear_square_evolve}(c-d).
Similar to the honeycomb dynamics above, the edge modes whose Floquet exponents reside in the Floquet edge gap are observed to pivot back-and-forth among the $b_{2m,0}$ and $a_{2m+1,1}$ lattice sites. {The energy is observed to oscillate with the same frequency as the lattice, $T = 1.5$}. There is an additional submode excited in Fig.~\ref{linear_square_evolve}(c) that is moving left-to-right and corresponds to the mode located near $\alpha(\pi/10) \approx 4$ in Fig.~\ref{linear_square_bands}(d). {Moreover, this right moving mode is {\it not} a protected solution i.e. it will scatter at defects.} {It was numerically verified that the left-moving mode interacted with a defect like the topologically protected mode shown in the top row of Fig.~\ref{honey_defect_intense_snap}{; i.e. this mode travelled through the defect}. On the other hand, the right-traveling mode was found to scatter off a defect like the mode displayed in the bottom row of Fig.~\ref{honey_defect_intense_snap}.} {For the cases we studied  it was} {unusual to have a system that simultaneously supports both topological and non-topological edge states.}

In Figs.~\ref{linear_square_evolve}(e-f) {which corresponds to Fig.~\ref{linear_square_bands}(f)} the $\pi$-offset edge profile is found to oscillate back-and-forth between the $a$ and $b$ sites with the same period as that of the helix ($T = 1.5$). This dynamics pattern resembles the $\pi$-offset honeycomb evolution shown in Fig.~\ref{linear_honeycomb_evolution}(e-f). {Between the honeycomb and square evolutions we can see that similar lattice {rotation patterns} {often} yield similar edge mode evolutions,} { when the underlying band structures are similar.}


\section{Edge Solitons}
\label{Nonlinear_modes}

In this section we explore {the} nonlinear ($\sigma \not=0$) equations given in Sec.~\ref{TBA_derive}.
Using an asymptotic analysis, valid in the {small} $\epsilon$ regime, we find true {\it edge soliton} solutions, i.e. edge modes modulated by a slowly-varying envelope function that satisfies the NLS equation. Direct numerical simulations are used {to} 
validate our asymptotic results. On the other hand, this asymptotic theory does not describe every possible edge mode found in Sec.~\ref{linear_theory}. As we {have remarked}, modes located and the Floquet edge can disappear as $\epsilon \rightarrow 0,$ and therefore are outside the scope of our {small $\epsilon$} analysis.

To leading order, we find 2D zig-zag edge mode solutions to Eq.~(\ref{honey_TBA_eq1})-(\ref{honey_TBA_eq2}) of the form
\begin{equation}
\label{asym_solns}
a_{mn}(z) = 0 \; , ~~~ b_{mn}(z) = C(y,z) b_n(\omega_0) e^{i m \omega_0} \; ,
\end{equation}
where the envelope function, $C(y,z)$, satisfies the NLS-type equation
\begin{align}
\nonumber
i \frac{\partial C}{\partial z} &- \alpha_0 C + i \alpha_0' C_y + \frac{ \alpha_0''}{2} C_{yy} - i\frac{  \alpha_0'''}{6} C_{yyy}+ \dots \\ \label{envelope_NLS_general}
& + \alpha_{\rm nl}(\omega_0) |C|^2 C + \dots = 0 \; ,
\end{align}
such that $\tilde{\alpha}_0^{(j)} = \frac{d^j \tilde{\alpha}}{d \omega^j} \big|_{\omega = \omega_0}$ and the function $C(y,z)$ varies slowly in $y$, i.e. {$|\partial_yC| \ll 1$}. {The variable $y$ is slowly-varying in the $m$-direction.} The details of the analysis are given in Appendix \ref{Asymptotic_Analysis}.
In the direction perpendicular to the zig-zag edge, the solution in (\ref{asym_solns}) decays like the stationary mode $b_n^s= r^n$ for $|r| < 1$ [the value of $r(\omega_0)$ is defined in Eq.~(\ref{define_r})]. This solution is derived under a narrow-band approximation, which assumes that only the frequencies, $\omega$, near  $\omega_0$ make substantial contributions to the solution.
Then by setting 
$C(y,Z) = \tilde{C}(y,Z)e^{-i \alpha_0 Z}$ we obtain (to leading order) the NLS equation
\begin{equation}
\label{envelope_NLS}
i \frac{\partial \tilde{C}}{\partial z}  + i \alpha_0' C_y + \frac{ \alpha_0''}{2} \tilde{C}_{yy} + \alpha_{\rm nl}(\omega_0) |\tilde{C}|^2 \tilde{C}  = 0 \; ,
\end{equation}
which for $\alpha_0'' > 0$ (focusing) has the traveling soliton solution
\begin{equation}
\tilde{C}(y,z) = \mu \sqrt{ \frac{\alpha_0''}{\alpha_{\rm nl}} } {\rm sech}\left[ \mu (y - \alpha_0'z) \right] e^{ i  \frac{\mu^2 \alpha_0''}{2} z}  \; , 
\end{equation}
with $\mu \ge 0$.
At points of inflection [like at $\omega = \pi/2$ in Fig.~\ref{linear_honey_bands}(a)], $\alpha_0'' = 0$ and therefore the leading order dispersive term is now third-order and we get the higher-order NLS equation
\begin{equation}
\label{envelope_NLS_third}
i \frac{\partial \tilde{C}}{\partial Z}  + i \alpha_0' C_y  - i\frac{ \alpha_0'''}{6} \tilde{C}_{yyy} + \alpha_{\rm nl}(\omega_0) |\tilde{C}|^2 \tilde{C}  = 0 \; ,
\end{equation}
which does {\it not} support solitons.

Next, we numerically verify these asymptotic results by direct numerical simulations performed on the full tight-binding system (\ref{honey_TBA_eq1})-(\ref{honey_TBA_eq2}). Here we restrict our attention to the nonlinear modes found using our asymptotic theory (corresponding to red curves in Fig.~\ref{linear_honey_bands}). Analysis of the remaining modes is {outside the scope of this paper.}

{To} initialize the simulations we take functions of the form in Eq.~(\ref{asym_solns}). For the figures shown here the linear decaying mode, $b_n^s$, were computed numerically, but we did check that the asymptotic solution, $b_n^s = r^n$, gave similar results. We normalize the edge mode two-norm so that $<b_n, b_n > = 1.$ The envelope is initialized by the localized function
\begin{equation}
C(y =  \mu m,z) = A~ {\rm sech}\left( \mu m \right) \; , ~~~ A \ge 0 \; ,
\end{equation}
for an amplitude, $A$, chosen to balance the leading-order dispersion term (either $\alpha_0'' \mu^2$ or $\alpha_0''' \mu^3$) with the cubic nonlinearity, $A^2 \alpha_{\rm nl}.$

\begin{figure} [h]
\includegraphics[scale=.41]{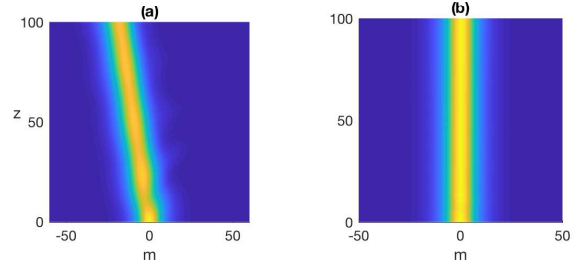}
\caption{Evolution of nonlinear edge magnitude, $|b_{m,0}(z)|$, found by solving (\ref{honey_TBA_eq1})-(\ref{honey_TBA_eq2}) using initial condition (\ref{asym_solns}). The parameters in (a) and (b) are the same as those in Figs.~\ref{linear_honey_bands}(a) and \ref{linear_honey_bands}(b), respectively, except $\sigma = \epsilon, \mu = 0.2,$ and (a) $\omega_0 = \pi/2, \alpha_0' = -0.1900, \alpha_0'' = 0, \alpha_0''' = 0.7482, \alpha_{\rm nl} = 0.2382$ and (b) $\omega_0 = 3\pi/4, \alpha_0' = 0, \alpha_0'' = 0.3622, \alpha_0''' = 0, \alpha_{\rm nl} = 0.1716.$}
\label{nl_honey_evolve}
\end{figure}

\begin{figure} [h]
\includegraphics[scale=.55]{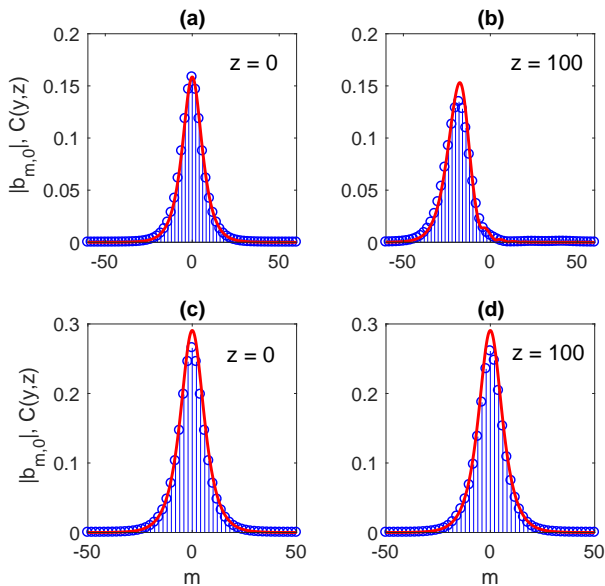}
\caption{(Color online) Profile comparison between the discrete solution (blue stems), $b_{m,0}(z)$, and envelope (red curve), $C(y,z)$. The parameters in panels (a-b) are the same as those in Fig.~\ref{nl_honey_evolve}(a), and similarly for panels (c-d) and Fig.~\ref{nl_honey_evolve}(b).}
\label{nonlinear_profiles}
\end{figure}

Two edge mode evolutions are displayed in Fig.~\ref{nl_honey_evolve}. In the first case, a traveling mode whose corresponding linear edge mode [located at $\omega_0 = \pi/2$ in Fig.~\ref{linear_honey_bands}(a)] is at a point of inflection ($\alpha_0'' = 0$). This implies that the envelope is governed by the higher-order NLS Eq.~(\ref{envelope_NLS_third}) {which}  
does not support {pure} soliton modes. A closer comparison of the full discrete solution on the domain boundary, $b_{m0}$, and the envelope $C(y,z)$ is shown in Fig.~\ref{nonlinear_profiles}, where the initial and final solution profiles are shown in Figs.~\ref{nonlinear_profiles}(a) and \ref{nonlinear_profiles}(b), respectively. The envelope captures the translation and a small dispersive tail which becomes more pronounced as $z$ grows.

The next case we consider is when the envelope function satisfies the classic NLS equation (\ref{envelope_NLS}). We choose the same lattice parameters that correspond to the linear Floquet bands shown in Fig.~\ref{linear_honey_bands}(b). At the point $\omega_0 = 3 \pi /4$ the value of $\alpha_0'' $ is positive indicating that the NLS equation (\ref{envelope_NLS}) is focusing and thus admits soliton modes. The evolution of such a nonlinear edge mode is shown in Fig.~\ref{nl_honey_evolve}(b). The mode is stationary because $\alpha_0' = 0$. A closer look at the initial and final edge soliton profiles is given in Figs.~\ref{nonlinear_profiles}(c-d). Comparing these two figures we see that the final magnitude is nearly the same as that which was initially injected.
In both this case and previous one there is a small gap between the envelope and discrete mode peaks that may be attributed to some energy being transferred from the $b_{mn}$ mode into the $a_{mn}$ mode. 
{We also remark that} in the latter case it so happens that $|b_0| < 1$, so there is also an initial gap at $z = 0.$

\begin{figure} [h]
\includegraphics[scale=.44]{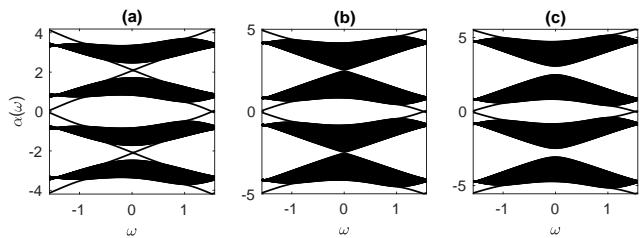}
\caption{Floquet bands for a $\pi$-offset honeycomb lattice rotation. The value of $\epsilon$ is (a) $0.75/\pi\approx 0.239$, (b) $0.2$, and (c) $0.18$. The other parameters are the same as those in Fig.~\ref{linear_honey_bands}(f).}
\label{finite_eps_evolve}
\end{figure}

The asymptotic results shown here only cover edge modes located near the Floquet center.
Moreover, the curves at the Floquet edges in Figs.~\ref{linear_honey_bands} and \ref{linear_square_bands} are the result of finite $\epsilon$ and disappear as $\epsilon \rightarrow 0 .$
This is highlighted in Fig.~\ref{finite_eps_evolve}. 
In this case there are two distinct families of edge modes: one that bifurcates from the stationary mode [located near $\alpha = 0$ in Fig.~\ref{finite_eps_evolve}(a)] and another [located near {$\alpha = \pm \pi/T$} in Fig.~\ref{finite_eps_evolve}(a)] {separate one that arises} when the helix period is not  {necessarily small} in comparison to the characteristic propagation distance scale {(not oscillating too fast)}.
As $\epsilon$ approaches zero the edge modes at the Floquet edge are observed to disappear, leaving {only} those modes predicted by the asymptotic theory in Appendix~\ref{Asymptotic_Analysis}. Indeed, to understand a mode that oscillates like the linear edge state in Fig.~\ref{linear_honeycomb_evolution}(e) requires a true coupled{-mode} theory, i.e. not setting $a_{mn} = 0$, to account for this back-and-forth energy transfer. To understand this finite $\epsilon$ edge modes is beyond the scope of this paper.

\section{Lattice Defects}
\label{defect_sec}

Lattice defects and imperfections are common in any real photonic lattice. 
Here we consider a scenario where many lattice sites are absent along the boundary, thereby introducing a boundary notch, or wall. The absence of any lattice sites along the boundary means the wave field is effectively zero there.
To implement these boundary effects into the tight-binding model above we set $a_{mn} = b_{mn} = 0$ at the lattice defect locations.
To track the evolution dynamics of the edge mode as it confronts the defect two quantities are monitored: the maximum intensities, $|a_{mn}(z)|^2$ and $|b_{mn}(z)|^2$, and the {`participation'} ratio 
\begin{equation}
\label{part_number}
P_b(z) = \frac{\left( \sum_{m,n} |b_{mn}|^2\right)^2}{ \sum_{m,n} |b_{mn}|^4 } \; .
\end{equation}
The first quantity measures the edge mode peaks, while the latter gives a measure of the pulse width since (\ref{part_number}) is 
{p}roportional to it.

\begin{figure} [h]
\includegraphics[scale=.29]{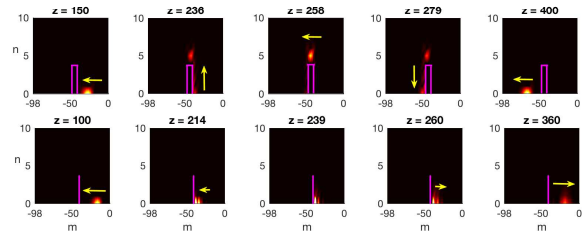}
\caption{Intensity snapshots, $|b_{mn}(z)|^2$, for a (top row) topologically protected mode and (bottom row) non-topologically protected mode. The defect barrier is located in the region $[-46, -40] \times [0 , 4].$}
\label{honey_defect_intense_snap}
\end{figure}

\begin{figure} [h]
\includegraphics[scale=.65]{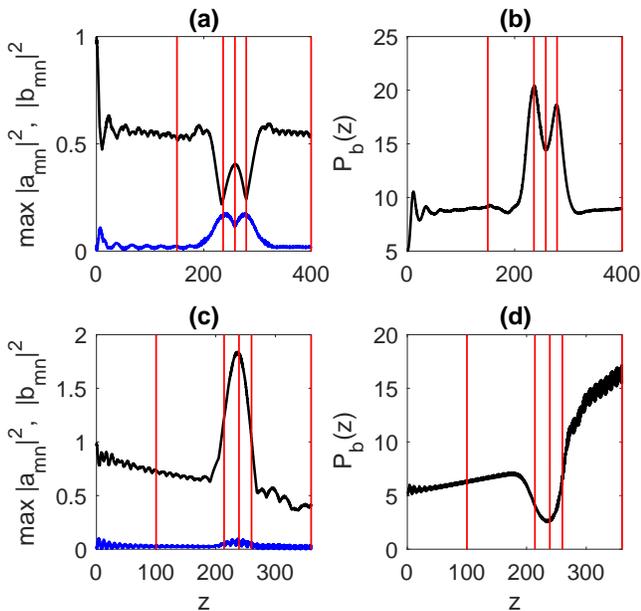}
\caption{(Color online) Maximum intensity evolution of $|a_{mn}(z)|^2$ (blue curve) and $|b_{mn}(z)|^2$ (black curve) for the (a) topologically and (c) non-topologically protected modes shown in Fig.~\ref{honey_defect_intense_snap}. The corresponding participation ratios (\ref{part_number}) are displayed in panels (b) and (d), respectively.
Vertical red lines correspond to the snapshots in Fig.~\ref{honey_defect_intense_snap}.}
\label{honey_defect_stats}
\end{figure}

We focus our attention on {three} illuminating examples. In the first situation we evolve a linear {($\sigma = 0$)} topologically protected edge mode whose corresponding Floquet exponent is given in Fig.~\ref{linear_honey_bands}(a) at $\omega_0 = \pi/2.$ The evolution of the edge mode when it comes into contact with the lattice defect is shown in Fig.~\ref{honey_defect_intense_snap}. In most non-topological systems significant {scattering} would be expected. Here, however, the edge mode tracks around the boundary notch and exhibits no backscattering. 
The intensity and participation ratio evolutions in Fig.~\ref{honey_defect_stats} shed some light on the edge mode-defect dynamics.
Along the defect boundaries perpendicular to the $n=0$ axis the boundaries have armchair configurations.
It is here that the energy begins to evenly distribute between the modes [see Fig.~\ref{honey_defect_stats}(a)]. In addition, the outer edge mode, $b_{m0}$, is found to spread to many sites (become wider) [see Fig.~\ref{honey_defect_stats}(b)] along these armchair boundaries.
The outgoing intensity is observed to have nearly the same magnitude as the incoming value.

The second case we consider is that of a non-topologically protected mode. The corresponding Floquet exponent is shown in Fig.~\ref{linear_honey_bands}(b) at $\omega_0 = \pi/2.$ Recall that this mode resulted from elongating the waveguide geometries in the direction perpendicular to the zig-zag boundary.
The edge mode dynamics shown in Fig.~\ref{honey_defect_intense_snap} contrast from those in the previous case. Upon making contact with the defect boundary the edge mode {scatters} back in the opposite direction. {From Fig.~\ref{honey_defect_stats}(c) the mode intensity is found to double at the defect corner, meanwhile Fig.~\ref{honey_defect_stats}(d) shows that the pulse width is cut in half ({becomes} more localized).} Upon reflecting backwards significant dispersion in the edge state is observed.

{Finally, we consider the interaction of a nonlinear edge mode with a defect. In particular, we focus on states we refer to as {\it topologically protected edge solitons}. These modes are characterized by linear Floquet bands with sign-invariant group velocity [see Fig.~\ref{linear_honey_bands}(a)] whose envelopes, to leading order, satisfy the NLS equation (\ref{envelope_NLS}). In general, we expect these states to combine the robustness of solitons with the unidirectionality of a protected state. Consider dispersion bands like those shown in Fig.~\ref{linear_honey_bands}(a). At the point $\omega = 5 \pi/8 \approx 1.963$ we have $\alpha_0'' > 0 $ and $|\alpha_0'''| \ll 1$, yet $\alpha_0' <0.$ In this case the traveling envelope satisfies the focusing NLS equation and therefore it is a true soliton. One such case is {displayed} in Fig.~\ref{nl_defect_evolve}. This nonlinear edge mode is observed to track around the defect just like the linear protected mode in Fig.~\ref{honey_defect_intense_snap}. The outgoing soliton intensity is found to be nearly the same as the incoming magnitude. The maximum intensity and the participation ratio for this case are shown in Fig.~\ref{nl_defect_stats}. These values closely resemble the linear ones in Figs.~\ref{honey_defect_stats} (a-b). One difference is that some loss is in the magnitude is observed in nonlinear case. This may be attributed to the fact that we are not in the pure NLS (\ref{envelope_NLS}), but instead of a small amount of higher-order dispersion, since $\alpha_0''' \not=0$. This extra dispersion results in a small amount of radiation being emitted from the mode {Decreasing the value of $\mu$ should improve the asymptotic NLS approximation}.

\begin{figure} [h]
\includegraphics[scale=.27]{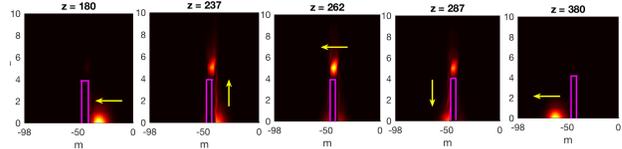}
\caption{Intensity snapshots, $|b_{mn}(z)|^2$, for a topologically protected edge soliton ($\sigma = \epsilon$). The defect barrier is located in the region $ [-46, -40] \times [0 , 4].$}
\label{nl_defect_evolve}
\end{figure}

\begin{figure} [h]
\includegraphics[scale=.45]{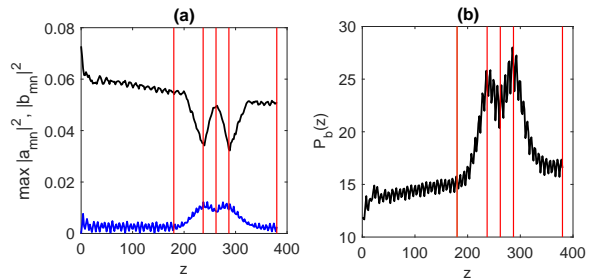}
\caption{(Color online) (a) Maximum intensity evolution of $|a_{mn}(z)|^2$ (blue curve) and $|b_{mn}(z)|^2$ (black curve) for the topologically protected edge soliton shown in Fig.~\ref{nl_defect_evolve}. (b) The corresponding participation ratio (\ref{part_number}).
Vertical red lines correspond to the snapshots in Fig.~\ref{nl_defect_evolve}.}
\label{nl_defect_stats}
\end{figure}
}

\section{Conclusions}
\label{conclude}

We have introduced a direct method for deriving tight-binding approximations of beam propagation in  {general 2D longitudinally} driven 
waveguide arrays.  {As prototypes we examined periodically} driven honeycomb and staggered square lattices. The new set of governing equations we derived allows us to find and examine unexplored edge modes {with} rather complicated sublattice rotation patterns such as: counter rotation, different radii, different frequency{, different structure} and $\pi$-phase offset. We computed the linear Floquet bands for these rotation configurations and also examined their evolutions. 

An asymptotic theory {was developed; it} showed that the nonlinear edge modes behave as linear edge states modulated by a slowly-varying envelope that satisfies the 1D nonlinear Schr\"odinger equation. {Ideal nonlinear edge modes} combine the scatter-free topological properties of the linear problem with the robust nature of solitons.


Finally, topologically protected modes{, both linear and nonlinear,} were shown to possess unidirectionality (absence of backscatter), even when encountering {strong} lattice defects. In the absence of topological protection the mode simply reflects off a barrier. The approach for deriving tight-binding equations and analyzing photonic topological insulators {allows the exploration of many new Floquet {and non-Floquet} lattice systems.}

\appendix

{
\section{Additional Tight-binding Details}
\label{TBA_details}

Here we {provide} some additional comments{/details} on deriving the honeycomb lattice tight-binding approximation in Sec.~\ref{TBA_derive}. The derivation for the staggered square lattice follows in a similar manner.
The ansatz in Eq.~(\ref{ansatz_define}) is defined in terms of the two Gaussian orbital functions $\widehat{\phi}_{1,{\bf v}}({\bf r},z)$ and $\widehat{\phi}_{2,{\bf v}} ({\bf r},z)$ centered at the points ${\bf v}$ and ${\bf v} + {\bf d} + \Delta {\bf h}_{12}(z)$, respectively.  These evanescent functions are taken to satisfy a linear ($\sigma = 0$) and local (${\bf r} \rightarrow {\bf r}_j, {\bf r}_k$) set of weakly-rotating ($|{\bf A}_z| \ll 1$) equations
\begin{align}
\label{local_eqn1}
& i \frac{\partial \widehat{\phi}_{1,{\bf v}}}{\partial z} + \nabla^2 \widehat{\phi}_{1,{\bf v}} - V_{\rm loc}({\bf r} - {\bf{v}}) = 0 \; , \\
\label{local_eqn2}
& i \frac{\partial \widehat{\phi}_{2,{\bf v}}}{\partial z} + \nabla^2 \widehat{\phi}_{2,{\bf v}} - V_{\rm loc}({\bf r} - ({\bf{d}}+{\bf{v}}) - \Delta {\bf h}_{21}(z)) = 0 \; ,
\end{align}
where 
\begin{equation}
V_{\rm loc}({\bf r} )= V_0^2 \left( \frac{x^2}{\sigma_x^2} + \frac{y^2}{\sigma_y^2} \right) \; .
\end{equation}
In particular, the normalized orbital functions used are
\begin{align}
& \widehat{\phi}_{1,{\bf v}}({\bf r} , z) = \phi_{1,{\bf v}}({\bf r}) e^{- i E z} 
\; , \\
& \widehat{\phi}_{2,{\bf v}}({\bf r} , z) = \phi_{2,{\bf v}}({\bf r},z) e^{- i E z} \; ,
\end{align}
where $\phi_{1,{\bf v}}({\bf r})  = \phi ({\bf r} - {\bf v}) $ and $ \phi_{2,{\bf v}} ({\bf r} , z)= \phi ({\bf r} - ({\bf d} + {\bf v}) - \Delta {\bf h}_{21}(z))$ for
\begin{equation}
\phi({\bf r}) = \sqrt{\frac{V_0}{\pi \sqrt{\sigma_x \sigma_y} } } e^{- \frac{V_0}{2} \left[ \frac{ x^2}{ \sigma_x} + \frac{ y^2}{ \sigma_y} \right] } \; , ~~
E = V_0 \left( \frac{1}{\sigma_x} + \frac{1}{\sigma_y} \right) \; .
\end{equation}
Substituting ansatz (\ref{ansatz_define}) into Eq.~(\ref{LSE_5}) for the Gaussian orbitals that satisfy Eqs.~(\ref{local_eqn1})-(\ref{local_eqn2}) yields 
\begin{align}
&\sum_{\bf v} \bigg[  \bigg( i \frac{d a_{\bf v}}{d z} \phi_{1,{\bf v}} +  i \frac{d b_{\bf v}}{d z} \phi_{2,{\bf v}}  \bigg) \\ \nonumber
&+  i V_0  \frac{ (\Delta {\bf h}'_{21}(z))_x [{\bf r} - ( {\bf v} + {\bf d}  )- \Delta {\bf h}_{21}(z)]_x}{\sigma_x}  b_{\bf v}  \phi_{2,{\bf v}}  \\ \nonumber
& + i V_0  \frac{ (\Delta {\bf h}'_{21}(z))_y [{\bf r} - ( {\bf v} + {\bf d}  )- \Delta {\bf h}_{21}(z)]_y}{\sigma_y} b_{\bf v}  \phi_{2,{\bf v}} \\ \nonumber
& + {\bf r} \cdot {\bf A}_{z} \big[ a_{\bf v} \phi_{1,{\bf v}} +  b_{\bf v} \phi_{2,{\bf v}} \big]  \\ \nonumber
& -  a_{\bf v} \phi_{1,{\bf v}} \left[V({\bf r},z) - V_{\rm loc}({\bf r}- {\bf v})  \right]   \\ \nonumber
& - b_{\bf v} \phi_{2,{\bf v}} \left[ V({\bf r},z) -   V_{\rm loc}({\bf r} - ({\bf v}+{\bf d} ) - \Delta {\bf h}_{21}(z)  ) \right]    \bigg] e^{i k \cdot {\bf v}}  \\ \nonumber 
& = 0 \; ,
\end{align}
where subscripts $x$ and $y$ denote the $\hat{\imath}$ and $\hat{\jmath}$ vector components, respectively.
We next multiply this equation by $\phi_{1,{\bf p}}({\bf r})$ and $\phi_{2,{\bf p}}({\bf r},z)$ and integrate to get the `a' and `b' mode equations given in Eqs.~(\ref{honey_TBA_eq1}) and (\ref{honey_TBA_eq2}), respectively. For deep lattices ($V_0 \gg1$) we may consider only the self and nearest-neighbor interactions to good approximation. 



Following integration of the orbital functions, several simplifications to the remaining equations can be made. 
All self-interaction phase terms can be removed by the simple transformation $a_{{\bf p} + {\bf v}}(z) = \tilde{a}_{{\bf p} + {\bf v}}(z)\exp(i \Phi_{{\bf p} + {\bf v}}(z))$ and $b_{{\bf p} + {\bf v}}(z) = \tilde{b}_{{\bf p} + {\bf v}}(z)\exp(i \Theta_{{\bf p} + {\bf v}}(z))$ where
\begin{align}
\nonumber
& \Phi_{{\bf p} + {\bf v}}(z) = ( {\bf p} +{\bf v}) \cdot {\bf A}(z) -  z \bigg[V_0^2 - \frac{V_0}{2} \left( \frac{1}{\sigma_x} + \frac{1}{\sigma_y}  \right) \\ 
&~~~~~~~~~~~~~~~~~~~~~~~~~~~~~ - V_0^3 \sqrt{\frac{\sigma_x \sigma_y}{(1 + \sigma_x V_0) (1 + \sigma_y V_0 )}}  \bigg] \; , \\ \nonumber
& \Theta_{{\bf p} + {\bf v}}(z) = ( {\bf p} +{\bf d } + {\bf v}) \cdot {\bf A}(z) +  \int_0^z \Delta {\bf h}_{21}(\zeta) \cdot {\bf A}_{\zeta}(\zeta) d\zeta \\ 
- & z \bigg[ V_0^2 - \frac{V_0}{2} \left( \frac{1}{\sigma_x} + \frac{1}{\sigma_y}  \right)  - V_0^3 \sqrt{\frac{\sigma_x \sigma_y}{(1 + \sigma_x V_0) (1 + \sigma_y V_0 )}}  \bigg] \; .
\end{align}

Next it is observed that $ \left| d a_{\bf p} / d z \right|=\left| d b_{\bf p} / d z \right|= \mathcal{O}\left( \exp[{- V_0 |{\bf d}|^2/ (4 \sigma)}] \right) $.
Since the off-diagonal derivative terms are of the form $\exp[{- V_0 |{\bf d}|^2/ (4 \sigma)}]  d a_{\bf p} / d z$ they are exponentially smaller than all other terms and may be neglected.
We {only} consider the remaining dominant terms. These terms are numerically identified to be those with a coefficients of $V_0$ or $V_0^2$ (since we are in the deep lattice limit) and all other terms are dropped. Finally, to simplify working with these equations we relabel our grid in terms of a two-dimensional square lattice indexed as in Fig.~\ref{honeycomb_fig}. }

\section{Tight-binding Approximation Coefficients}
\label{TBA_coefficients}

In this section the tight-binding approximation coefficients for the honeycomb lattice given in Eqs.~(\ref{honey_TBA_eq1})-(\ref{honey_TBA_eq2}),  and the staggered square lattice (\ref{square_TBA_eq1})-(\ref{square_TBA_eq2}). The subscripts $x$ and $y$ denote the $\hat{\imath}$ and $\hat{\jmath}$ vector components, respectively.
The nonlinearity coefficient in both cases is given by 
\begin{equation}
\sigma = \frac{\gamma  V_0}{2 \pi \sqrt{\sigma_x \sigma_y}} \; .
\end{equation}

\subsection{Honeycomb Lattice}
\label{TBA_coefficients_honey}
The terms composing the honeycomb linear operator defined 
{below Eqs.~(\ref{honey_TBA_eq1})-(\ref{honey_TBA_eq2}) are}
\begin{align*}
\label{L_define}
\mathbb{L}_j(z)= &V_0^3  \sqrt{\frac{\sigma_x \sigma_y}{(1 + \sigma_x V_0) (1 + \sigma_y V_0 )}} \\
&  \times \bigg\{ 2 e^{- \frac{V_0}{4} \left[ \frac{[ {\bf d} - {\bf v}_j + \Delta {\bf h}_{21}(z)]_x^2}{\sigma_x(1 + V_0 \sigma_x)} + \frac{[ {\bf d} - {\bf v}_j + \Delta {\bf h}_{21}(z)]_y^2}{\sigma_y (1 + V_0 \sigma_y)} \right]} - 1 \bigg\} \\
  +  \frac{V_0^2}{4}& \left\{ \frac{[ {\bf d} - {\bf v}_j + \Delta {\bf h}_{21}(z)]^2_x}{\sigma_x^2} + \frac{[ {\bf d} - {\bf v}_j + \Delta {\bf h}_{21}(z)]^2_y}{\sigma_y^2} \right\} \; ,
\end{align*}
\begin{align*}
\mathbb{R}_j(z) & = \frac{V_0}{2} \bigg[ \frac{ \left[ \Delta {\bf h}_{21}'(z) \cdot ({\bf d} -  {\bf v}_j  + \Delta {\bf h}_{21}(z))\right]_x}{\sigma_x} + \\ 
& ~~~~~~~~~~~~ +  \frac{ \left[ \Delta {\bf h}_{21}'(z) \cdot ({\bf d} -  {\bf v}_j   + \Delta {\bf h}_{21}(z))\right]_y}{\sigma_y} \bigg]  \; ,
\end{align*}
and
\begin{equation*}
c_j(z) = e^{- \frac{V_0}{4} \left[ \frac{[{\bf d} - {\bf v}_j + \Delta {\bf h}_{21}(z)]_x^2}{\sigma_x} + \frac{[{\bf d} - {\bf v}_j + \Delta {\bf h}_{21}(z)]_y^2}{\sigma_y} \right]} \; ,
\end{equation*}
where $\Delta {\bf h}_{21}(z) = {\bf h}_2(z) - {\bf h}_1(z),$ and the vectors ${\bf d} = (1,0), {\bf v}_0 = {\bf 0}, {\bf v}_1 = (3,\sqrt{3})/2$ and ${\bf v}_2 = (3,-\sqrt{3})/2$.

{In the special case of same rotation, same phase the following occur: $\Delta {\bf h}_{21}(z) = 0, \mathbb{R}_j(z) = 0,$ and the functions $c_j(z)$ and $\mathbb{L}_j(z) $ are constant for each $j$. Rescaling Eqs.~(\ref{honey_TBA_eq1})-(\ref{honey_TBA_eq2}) by $Z = c_0 \mathbb{L}_0 z$ gives the system considered in \cite{AbCuMa2014}, namely 
\begin{align}
& i \frac{d a_{mn}}{dZ} + e^{i {\bf d} \cdot {\bf A}(Z) } \left( \mathcal{L}_-(Z) b \right)_{mn} + \sigma |a_{mn}|^2 a_{mn} = 0 \; , \\
& i \frac{d b_{mn}}{dZ} + e^{-i {\bf d} \cdot {\bf A}(Z) } \left( \mathcal{L}_+(Z) a \right)_{mn} + \sigma |b_{mn}|^2 b_{mn} = 0 \; ,
\end{align}
where
\begin{align*}
 \left( \mathcal{L}_{-}(Z) b \right)_{mn} & =   b_{mn} \\
+  &  \rho \left[  b_{m-1,n-1} e^{- i \theta_1(Z)} + b_{m+1,n-1} e^{- i \theta_2(Z)}  \right]\; ,
\end{align*}
\begin{align*}
 \left( \mathcal{L}_{+}(Z) a \right)_{mn} & =   a_{mn} \\
+& \rho \left[ a_{m+1,n+1} e^{ i \theta_1(Z)} + a_{m-1,n+1} e^{ i \theta_2(Z)} \right] \; ,
\end{align*}
for the geometric parameter
\begin{equation}
\rho = \frac{c_1 \mathbb{L}_1}{c_0 \mathbb{L}_0} = \frac{c_2 \mathbb{L}_2}{c_0 \mathbb{L}_0}   \; .
\end{equation}
The anisotropic Floquet bands considered in Fig.~\ref{linear_honey_bands}(b) correspond to $\rho < 1/2$.
}

\subsection{Staggered Square Lattice}
\label{TBA_coefficients_square}
Here we give the definitions for the linear terms defined {below Eqs.~(\ref{square_TBA_eq1})-(\ref{square_TBA_eq2}) are}
\begin{align*}
\label{L_define}
\mathbb{L}_j(z) = &V_0^3  \sqrt{\frac{\sigma_x \sigma_y}{(1 + \sigma_x V_0) (1 + \sigma_y V_0 )}} \\
& ~~~~~~ \times \bigg\{ 2 e^{- \frac{V_0}{4} \left[ \frac{[  {\bf v}_j + \Delta {\bf h}_{21}(z)]_x^2}{\sigma_x(1 + V_0 \sigma_x)} + \frac{[  {\bf v}_j + \Delta {\bf h}_{21}(z)]_y^2}{\sigma_y (1 + V_0 \sigma_y)} \right]} - 1 \bigg\} \\
&  +  \frac{V_0^2}{4} \left( \frac{[  {\bf v}_j + \Delta {\bf h}_{21}(z)]^2_x}{\sigma_x^2} + \frac{[  {\bf v}_j + \Delta {\bf h}_{21}(z)]^2_y}{\sigma_y^2} \right) \; ,
\end{align*}
\begin{align*}
\mathbb{R}_j(z) & = \frac{V_0}{2} \bigg\{ \frac{ \left[ \Delta {\bf h}_{21}'(z) \cdot ( {\bf v}_j  + \Delta {\bf h}_{21}(z))\right]_x}{\sigma_x}  \\
& ~~~~~~~~~~+ \frac{ \left[ \Delta {\bf h}_{21}'(z) \cdot (  {\bf v}_j   + \Delta {\bf h}_{21}(z))\right]_y}{\sigma_y} \bigg\}  \; ,
\end{align*}
\begin{equation*}
c_j(z) = e^{- \frac{V_0}{4} \left[ \frac{[ {\bf v}_j + \Delta {\bf h}_{21}(z)]_x^2}{\sigma_x} + \frac{[ {\bf v}_j + \Delta {\bf h}_{21}(z)]_y^2}{\sigma_y} \right]} \; ,
\end{equation*}
for $j = \pm 1, \pm 2 $ and $\Delta {\bf h}_{21}(z) = {\bf h}_2(z) - {\bf h}_1(z)$. Here we {adopt} the convention ${\bf v}_{-j} \equiv - {\bf v}_j$.
The lattice vectors are defined by ${\bf v}_1 = (1,1) / \sqrt{2}$ and ${\bf v}_2 = (1,-1) / \sqrt{2}$. 

\section{Asymptotic Analysis}
\label{Asymptotic_Analysis}

Details for deriving asymptotic solution (\ref{asym_solns}) to the honeycomb lattice system (\ref{honey_TBA_eq1})-(\ref{honey_TBA_eq2}) are presented here. This analysis generalizes the calculation performed in \cite{AbCuMa2014} to cover {the  more general} non-synchronized rotation patterns {discussed in this paper}. 
The {periodic} functions $\Delta {\bf h}_{21}, {\bf \varphi}$ and ${\bf A}$ are all assumed to depend only on the fast variable $\zeta = z / \epsilon$, where $|\epsilon| \ll 1$ {and} weak nonlinearity of $\sigma = \epsilon \tilde{\sigma}$ is assumed.
To begin, we take functions of the form given in Eq.~(\ref{reduce_soln_dim}) and then expand
\begin{equation}
\label{expansions}
a_n = \sum_{j=0}^{\infty} \epsilon^j a_n^{(j)}(z,\zeta) \; , ~~~ b_n = \sum_{j=0}^{\infty} \epsilon^j b_n^{(j)}(z,\zeta) \; .
\end{equation}
For convenience, we gather all linear under the definitions $( \tilde{\mathcal{L}}_- b)_n \equiv   e^{i {\bf d} \cdot {\bf A}(\zeta) + i \varphi(\zeta)} \left( \mathcal{L}_-(\zeta) b \right)_{n} $ and $( \tilde{\mathcal{L}}_+ a)_n \equiv   e^{-i {\bf d} \cdot {\bf A}(\zeta) - i \varphi(\zeta)} \left( \mathcal{L}_+(\zeta) a \right)_{n}$.
Substituting expansions (\ref{expansions}) into Eqs.~(\ref{honey_TBA_eq1})-(\ref{honey_TBA_eq2}) and keeping the leading order terms gives 
\begin{equation}
\mathcal{O}\left(1/\epsilon\right): ~~~~  i \frac{\partial a_n^{(0)}}{\partial \zeta} = 0 \; , ~~~~ i \frac{\partial b_n^{(0)}}{\partial \zeta} = 0 \; ,
\end{equation}
which implies that $a_n^{(0)}(z,\zeta) = a_n^{(0)}(z)$ and $b_n^{(0)}(z,\zeta) = b_n^{(0)}(z)$.
At the next order we get
\begin{align}
\label{eqns_order_1}
\mathcal{O}(1): ~~~~ &  i \frac{\partial a_n^{(1)}}{\partial \zeta} = -  i \frac{d a_n^{(0)}}{d z}  - ( \tilde{\mathcal{L}}_- b^{(0)})_n  \; , \\ \nonumber
&   i \frac{\partial b_n^{(1)}}{\partial \zeta} = - i \frac{d b_n^{(0)}}{d z} - ( \tilde{\mathcal{L}}_+ a^{(0)})_n  \; .
\end{align}
To eliminate {secularities} these equations are rewritten as
\begin{align}
\label{secular_order_1}
&  i \frac{\partial a_n^{(1)}}{\partial \zeta} =  - \left[  ( \tilde{\mathcal{L}}_- b^{(0)})_n - \overline{( \tilde{\mathcal{L}}_- b^{(0)})}_n \right]  - f_- \; , \\ \nonumber
&   i \frac{\partial b_n^{(1)}}{\partial \zeta} = -\left[  ( \tilde{\mathcal{L}}_+ a^{(0)})_n - \overline{( \tilde{\mathcal{L}}_+ a^{(0)})}_n  \right] - f_+ \; ,
\end{align}
where
\begin{align}
&f_- = i \frac{\partial a_n^{(0)}}{\partial z} + \overline{( \tilde{\mathcal{L}}_- b^{(0)})}_n \; , \\ \nonumber
&f_+ = i \frac{\partial b_n^{(0)}}{\partial z} + \overline{( \tilde{\mathcal{L}}_+ a^{(0)})}_n \; ,
\end{align}
are zero at this order and {we define the average: $\overline{c} = \left( \int_0^T c(\zeta) d\zeta \right) /T,$ where $T$ is the lattice period}.  We consider solutions $a_n^{(0)} = 0$ which imply $b_n^{(1)} = 0$ (since it may be absorbed into $b_n^{(0)}$) and $ \overline{( \tilde{\mathcal{L}}_- b^{(0)})}_n = 0$. This latter equation gives solutions of the form $b_n^{(0)}(z)  = C(Z) b_n^s =  C(Z) r^n$, where $Z = \epsilon z$ and 
\begin{equation}
\label{define_r}
r = - \frac{\overline{\kappa_1 (\zeta, \omega) }}{\overline{\kappa_0(\zeta ,\omega) }} \; ,
\end{equation}
{where}
\begin{align*}
& \kappa_0 (\zeta ,\omega) =  e^{i {\bf d} \cdot {\bf A} + i \varphi}  c_0 \left( \mathbb{L}_0 - i \mathbb{R}_0 \right) \; , \\
& \kappa_1 (\zeta ,\omega) =  e^{i {\bf d} \cdot {\bf A} + i \varphi}  \big\{ c_1 \left( \mathbb{L}_1 - i \mathbb{R}_1 \right)e^{- i \omega - i \theta_1} \\
&~~~~~~~~~~~~~~~~~~~~~~~  + c_2 \left( \mathbb{L}_2 - i \mathbb{R}_2 \right)e^{i \omega - i \theta_2}  \big\} \; .
\end{align*}

The terms $f_{\pm}$ are expanded in series of $\epsilon$ and at the next order we get
\begin{align}
\mathcal{O}(\epsilon): ~~ &    i \frac{\partial b_n^{(2)}}{\partial \zeta} = - i \frac{d b_n^{(0)}}{d Z} - ( \tilde{\mathcal{L}}_+ a^{(1)})_n - \tilde{\sigma} |b_n^{(0)}|^2 b_n^{(0)} \; ,
\end{align}
where we have rescaled $Z = \epsilon z.$
Removing secularities, as we did in Eqs.~(\ref{eqns_order_1})-(\ref{secular_order_1}), yields
\begin{equation}
i \frac{d b_n^{(0)}}{d Z} + \overline{( \tilde{\mathcal{L}}_+ a^{(1)})_n} + \tilde{\sigma} |b_n^{(0)}|^2 b_n^{(0)} = 0 \; ,
\end{equation}
where Eq.~(\ref{secular_order_1}) gives $a_n^{(0)} = i \int_0^{\zeta}( \tilde{\mathcal{L}}_+(\zeta') b^{(0)})_n d\zeta'$.
Taking the inner product of this equation with the stationary mode $b_n^s$ gives the equation
\begin{equation}
\label{NLS_envelope}
i \frac{d C}{d Z} - \tilde{\alpha}(\omega) C + \alpha_{\rm nl}(\omega) |C|^2 C = 0 \; ,
\end{equation}
for the Floquet exponent 
\begin{equation}
\label{asym_floq_exp}
\tilde{\alpha}(\omega) = - \frac{i }{T } \int_0^T \int_0^{\zeta}  \mathcal{P}^*(\zeta; \omega)   \mathcal{N}(\zeta'; \omega) d\zeta' d\zeta \; , 
\end{equation}
defined in terms of the functions
\begin{align}
&\mathcal{P} = e^{i {\bf d} \cdot {\bf A} + i \varphi} \left[ c_0 \mathbb{L}_0 r + c_1 \mathbb{L}_1 e^{- i \omega - i \theta_1 } + c_2 \mathbb{L}_2 e^{ i \omega - i \theta_2}  \right] \; , \\ \nonumber
&\mathcal{N} = e^{i {\bf d} \cdot {\bf A} + i \varphi} \big[ c_0 \left( \mathbb{L}_0  - i \mathbb{R}_0 \right) r + \\ 
&~~~~~~~ c_1 \left( \mathbb{L}_1 - i \mathbb{R}_1 \right) e^{- i \omega - i \theta_1 } + c_2 \left( \mathbb{L}_2 - i \mathbb{R}_2 \right) e^{ i \omega - i \theta_2}  \big]  \; ,
\end{align}
and 
\begin{equation*}
\alpha_{\rm nl}(\omega) = \frac{<(b_n^s)^2,(b_n^s)^2>}{<b_n^s,b_n^s>} \tilde{\sigma} \; ,
\end{equation*}
using the inner product $<f_n,g_n> = \sum_n f_n^* g_n.$ 

We now convert the spectral NLS equation (\ref{NLS_envelope}) into {its corresponding} spatial version. To begin, we expand the Floquet exponent in a Taylor series expansion around the central frequency, $\omega = \omega_0$, by
\begin{equation}
\tilde{\alpha}(\omega) = \tilde{\alpha}_0 + \frac{(\omega - \omega_0)}{1!} \tilde{\alpha}_0' + \frac{(\omega - \omega_0)^2}{2!} \tilde{\alpha}_0'' + \cdots
\end{equation}
Next we take the inverse Fourier transform of this equation. When a narrow band is taken in $\omega$ this corresponds to a wide spatial profile. In other words, the spatial mode is slowly-varying along the zig-zag boundary, or in other words, it is nonzero at many lattice sites. Hence we take a Fourier transform that is continuous in space and replace the term $(\omega - \omega_0)$ with the derivative $ - i \partial_y.$ Doing this yields the nonlinear Schr\"odinger-type equation
\begin{align}
\nonumber
i \frac{\partial C}{\partial Z} &- \tilde{\alpha}_0 C + i \tilde{\alpha}_0' C_y + \frac{ \tilde{\alpha}_0''}{2} C_{yy} - i\frac{  \tilde{\alpha}_0'''}{6} C_{yyy}+ \dots \\ \label{envelope_NLS_general}
& + \alpha_{\rm nl}(\omega_0) |C|^2 C + \dots = 0 \; ,
\end{align}
where derivatives in $y$ are slowly-varying, i.e. $\left| \partial_y \right| \ll 1$. {If we consider the slowly varying length scale in $y$ to be $O(\nu)$ (or alternatively, calling the narrow band scale to be $O(\nu)$) then we can balance the weak nonlinearity by taking  $\tilde{\sigma} = O( \nu^p)$, where $p=2  \text{~if~} \alpha_0'' \neq 0$ or  $p=3 \text{~if~}  \alpha_0'' = 0, \alpha_0''' \neq 0$.}

\bibliography{PRA_TBA_Manuscript_8.3.17.bbl}

\end{document}